\documentclass{jfm}
\usepackage{natbib}
\usepackage{graphicx}
\usepackage{amsmath} 
\usepackage{amsfonts}
\usepackage{amssymb}
\usepackage{bm}
\usepackage{mathbbol}
\usepackage{xparse}
\usepackage{ifthen}	
\usepackage{xspace}

\usepackage{siunitx}
\usepackage{hyperref}
\usepackage{multirow}
\usepackage{array}
\usepackage{gensymb}
\usepackage{float}
\floatstyle{plain}
\restylefloat{figure}
\usepackage{subfig}

\newcommand{\svee}{{\hbox{\raise .4ex \hbox{${\scriptscriptstyle{\vee}}$}}}}
\newcommand{\swedge}{{\hbox{\raise .4ex \hbox{${\scriptscriptstyle{\wedge}}$}}}}

\newcommand{\CHI}{\hbox{\raise .4ex \hbox{$\chi$}}}

\newcommand{\scap}{\hbox{\raise .25ex
\hbox{${\operatornamewithlimits{\scriptstyle\bigcap}}$}}}
\newcommand{\scup}{\hbox{\raise .25ex
\hbox{${\operatornamewithlimits{\scriptstyle\bigcup}}$}}}

\newcommand{\deltacheck}
  {\overset {\lower .4ex \hbox{${\scriptscriptstyle{\hskip 2 pt\vee}}$}} \delta}

\newcommand{\Fcheck}
    {\overset {\lower .4ex \hbox{${\scriptscriptstyle{\hskip 2 pt\vee}}$}} F}

\newcommand{\fwedgehat}
    {\overset {\lower .6ex \hbox{${\scriptscriptstyle{\hskip 3 pt\wedge}}$}} f}
\newcommand{\fveecheck}
    {\overset {\lower .4ex \hbox{${\scriptscriptstyle{\hskip 2 pt\vee}}$}} f}
\newcommand{\fkcheck}
   {\overset {\lower .4ex \hbox{${\scriptscriptstyle{\hskip 1 pt\vee}}$}} {f_k}}

\newcommand{\raiseprime}{\hbox{\raise .3ex \hbox{${\scriptstyle{\prime}}$}}}

\newcommand{\Gcheck}
    {\overset {\lower .4ex \hbox{${\scriptscriptstyle{\hskip 2 pt\vee}}$}} G}

\newcommand{\gveecheck}
    {\overset {\lower .4ex \hbox{${\scriptscriptstyle{\hskip 2 pt\vee}}$}} g}

\newcommand{\hcheck}
    {\overset {\lower .4ex \hbox{${\scriptscriptstyle{\hskip 2 pt\vee}}$}} h}

\newcommand{\Kcheck}
  {\overset {\lower .4ex \hbox{${\scriptscriptstyle{\hskip 2 pt\vee}}$}} K}

\newcommand{\mucheck}
    {\overset {\lower .4ex \hbox{${\scriptscriptstyle{\hskip 2 pt\vee}}$}} \mu}

\newcommand{\norm}[1]{\|#1\|}

\newcommand{\varphicheck}
    {\overset {\lower .4ex \hbox{${\scriptscriptstyle{\hskip 1 pt\vee}}$}}
              \varphi}

\newcommand{\R}{\mathbb{R}}

\newcommand{\Rbar}
  {{\overset {\hskip -0.9 pt \lower 1.5pt \hbox{{\rule{6.7pt}{0.45pt}}}} \R}}
\newcommand{\subRbar}
   {{\overset {\hskip -0.8 pt \lower 1.5pt \hbox{{\rule{4.5pt}{0.5pt}}}} \R}}

\newcommand{\sign}{{\mathrm{sign}}}

\newcommand{\zeroveecheck}
    {\overset {\lower .4ex \hbox{${\scriptscriptstyle{\hskip 0.5 pt\vee}}$}} 0}

\newcommand{\mnorm}[1]{{\left\vert\kern-0.25ex\left\vert\kern-0.25ex\left\vert #1 \right\vert\kern-0.25ex\right\vert\kern-0.25ex\right\vert}}

\newcommand{\pd}[3]{\frac{\partial^{#3} #1}{\partial {#2}^{#3}}}

\newcommand{\eb}{\begin{equation}\begin{aligned}}
\newcommand{\en}{\end{aligned}\end{equation}}

\renewcommand{\theenumi}{\alph{enumi}}

\usepackage[usenames,dvipsnames]{xcolor}

\shorttitle{Exact Coherent Structures in Two-Dimensional Turbulence}
\shortauthor{D. Zhigunov and R. O. Grigoriev}

\title{Exact Coherent Structures in Fully Developed Two-Dimensional Turbulence}

\author{Dmitriy Zhigunov
\and Roman O. Grigoriev
\corresp{\email{roman.grigoriev@physics.gatech.edu}}}

\affiliation{School of Physics, Georgia Institute of Technology, 
Atlanta, GA 30332, USA}

\begin{document}

\maketitle

\begin{abstract}
This paper reports several new classes of unstable recurrent solutions of the {two-dimensional Euler equation} on a square domain with periodic boundary conditions. {These solutions are in many ways analogous to recurrent solutions of the Navier-Stokes equation which are often referred to as exact coherent structures. In particular, we find that recurrent solutions of the Euler equation are dynamically relevant: they faithfully reproduce large-scale flows in simulations of turbulence at very high Reynolds numbers. On the other hand, these solutions have a number of properties which distinguish them from their Navier-Stokes counterparts.} First of all, recurrent solutions of the Euler equation come in infinite-dimensional continuous families. Second, solutions of different types are connected, e.g., an equilibrium can be smoothly continued to a traveling wave or a time-periodic state. Third, and most important, {they are only weakly unstable and, as a result, fully-developed turbulence mimics some of these solutions} remarkably frequently and over unexpectedly long temporal intervals. 
\end{abstract}

\section{Introduction}

The important role of coherent structures in fluid turbulence has been both widely accepted and hotly debated for several decades. However, many open questions still remain. A quote from the seminal paper on this topic by \citet{hussain1986} sets the stage for the present study: ``The interaction between coherent structures and incoherent turbulence is the most critical and least understood aspect of turbulent shear flows. This coupling appears to be rather different from the classical notion of cascade; even considering the large and fine scales, they are not decoupled as widely presumed. The coupling can be intricate and of different kinds...'' Indeed, this pretty much sums up the state of knowledge to the present day.

We will follow Hussain in distinguishing coherent {\it structures} as those of size comparable to the transverse length scale of the shear flow as opposed to coherent {\it substructures} whose characteristic size corresponds to the  Taylor microscale.
These scales are only clearly distinguished at high Reynolds numbers ($Re$) where coherent structures can be considered inviscid. Rather interestingly, the greatest progress in understanding coherent structures to date has been made in the context of weakly turbulent flows where this scale separation disappears. Advanced numerical methods such as Newton-Krylov solvers \citep{Viswanath2007} enabled computation of unstable, recurrent (e.g., steady. time-periodic or {relative}-periodic) solutions of the governing equations in a variety of canonical shear flows. Many of such numerical solutions were found to have spatiotemporal structure similar to that of familiar coherent structures such as streamwise vortices and velocity streaks and hairpin vortices near a wall \citep{waleffe1998,waleffe2001,gibson2008,itano2009,shekar2018}; consequently, they have been termed exact coherent structures (ECSs). 
Recent numerical and experimental studies demonstrated that ECSs do not merely resemble turbulent flows, they also organize and guide the dynamics of {weak} turbulence in both two \citep{suri2018,suri2020} and three spatial dimensions \citep{krygier2021,crowley2022}. 

ECSs have already generated significant insight into weakly turbulent flows \citep{kawahara2012,graham2021}. Most notably, ECSs are found to capture self-sustaining processes that maintain wall-bounded turbulence \citep{Waleffe1997,Hall1991}. ECSs also elucidate the transition from laminar flow to turbulence explaining both the ``bypass'' mechanism in linearly stable flows  \citep{khapko2016} and the formation of chaotic sets underpinning sustained turbulence \citep{kreilos2012}. Despite these successes, due to the lack of scale separation in {weakly turbulent} flows, it remains unclear how much of this understanding carries over to fully developed turbulence. 

Extending the ECS framework to higher $Re$ proved challenging due to both conceptual and technical challenges. As $Re$ increases, the range of scales accessible to turbulence becomes larger, and the number of distinct ECSs grows very quickly. Furthermore, each ECS becomes more unstable. And, to complicate things even further, it becomes even more expensive to find ECSs through direct numerical simulations (DNS) which effectively become intractable at high $Re$ due to the high spatial and temporal resolution requirements. A key conceptual challenge is related to the limit $Re\to\infty$. The Euler equation is a singular limit of the Navier-Stokes equation, which makes it difficult to establish a relation between dynamically relevant recurrent solutions of the two equations {beyond some relatively loose, although important, constraints \citep{batchelor1956,okamoto1994,gallet2013}}.

One of the most notable successes in continuing recurrent solutions to higher $Re$ is the computation of attached eddies { \citep{eckhardt2018,yang2019,doohan2019,azimi2020} and their bulk analogues \citep{deguchi2015, eckhardt2018} in a variety of canonical 3D shear flows}. In wall units, these solutions become independent of $Re$, just like the wall-bounded fully developed turbulent flows. It is important to note that these solutions represent coherent {\it substructures}. To the best of our knowledge, no examples of ECS have been found in high-$Re$  {3D turbulent flows at the scale comparable to either the system size (e.g., distance between the boundaries) or the scale of the forcing; such ECSs correspond to recurrent solutions of the Euler equation. It should be pointed out, however, that, some -- not necessarily dynamically relevant -- steady and time-periodic solutions representing large-scale flows at fairly high $Re$ have been computed through continuation in 2D \citep{kim2010,kim2015,kim2017} and 3D \citep{wang2007}.  In principle, recurrent large-scale flows could also be computed with the help of large eddy simulations \citep{rawat2015, hwang2016}, although such flows do not represent formal solutions of either the Navier-Stokes or Euler equation.}

The objective of this paper is therefore to introduce such large-scale recurrent solutions of the incompressible Euler equation in two spatial dimensions, where fully resolved DNS of turbulent flow can be performed for relatively high $Re$. There is a rich history of computing solutions of the Euler equation using various analytical tools. Even leaving singular solutions involving point vortices aside, quite a few examples of absolute and relative equilibria have been found. They include isolated circular and elliptic vortices \citep{lamb1924}, translating pairs of counter-rotating vortices \citep{pierrehumbert1980,saffman1982}, rotating arrays of two or more vortices \citep{dritschel1985,carton1989,crowdy2002rw}, and stationary multipolar vortex arrays \citep{crowdy1999,crowdy2002eq,tur2004,xue2017}.

This paper reports several new classes of recurrent solutions -- equilibria, traveling waves, and time-period{ic} states -- of the 2D Euler equation on a domain with periodic boundary conditions which can be thought of as generalizations of vortex crystals \citep{aref2003}. A distinguishing feature of these solutions is their dynamical relevance for high-$Re$ statistically stationary turbulent flows which balance forcing and dissipation: just like in the case of {weakly turbulent} flows, we find recurrent solutions to organize and guide the dynamics of turbulent flow on large scales. As expected, the library of inviscid recurrent solutions is found to be far larger than that describing viscous  flows: unlike their analogues for {weakly turbulent} flows which are isolated, solutions of the Euler equation are found to belong to continuous families spanned by an infinite number of parameters. 

The rest of the paper is organized as follows. Section \ref{sec:desc} describes the problem setup investigated here and Section \ref{sec:limit} discusses the relation between solutions of the Euler and Navier-Stokes equation in the high-$Re$ limit. Our results are presented in Section \ref{sec:results}, their implications for the problem of fully developed fluid turbulence are discussed in Section \ref{sec:disc}, and conclusions are presented in Section \ref{sec:conc}.

\section{Problem description}\label{sec:desc}

We consider an incompressible Newtonian fluid in two spatial dimensions. In the presence of forcing and dissipation, its dynamics are governed by the Navier-Stokes equation, \eb\label{eq:nseu}
\partial_t{\bf u}+ \left({\bf u}\cdot\nabla\right){\bf u} = -\nabla p+Re^{-1}\nabla^2{\bf u}+{\bf f},
\en 
where $p$ is the pressure, ${\bf f}$ represents the external forcing, and $Re$ is the Reynolds number. In two spatial dimensions, it can be conveniently rewritten in terms of vorticity $\omega=\partial_xu_y-\partial_yu_x=-\nabla^2\psi$,
\eb\label{eq:nse}
\partial_t\omega+ \left({\bf u}\cdot\nabla\right)\omega = Re^{-1}\nabla^2\omega+\varphi,
\en
where $u_x=\partial_y\psi$ and $u_y=-\partial_x\psi$ are the components of the velocity field, $\psi$ is the stream function, and $\varphi=\partial_xf_y-\partial_yf_x$. In this study, we confine our attention to spatially-periodic domains $0\le x,y<2\pi$ with stationary checkerboard forcing 
\eb
\varphi=\sin(Nx)\sin(Ny),
\en
where $N=4$. With this choice of nondimensionalization, the length, time, velocity, and vorticity scales are all $O(1)$. 

The frequency of the spatial forcing, $k_f=\sqrt{2}N\approx 6$ is chosen to be reasonably well separated from both the dominant wavenumber $k_0=1/\sqrt{2}$ describing the large-scale flow on the low wavenumber side and the inertial range on the higher wavenumber side. This choice of the forcing frequency ensures that the direct (enstrophy) cascade takes place over most of the wavenumber range resolved in the simulations while the inverse (energy) cascade is constrained to the narrow range $[k_0,k_f]$. As far as the structure of the large-scale flows at high $Re$ is concerned, the particular choice of the forcing profile does not appear to play a noticeable role according to both our simulations and the results of previous systematic studies \citep{gallet2013,kim2017}. Note that, unlike, say, a Kolmogorov forcing profile \citep{tithof2017}, a checkerboard profile breaks continuous translational symmetry in both spatial directions, constraining the types of ECSs that can be found at moderate $Re$ \citep{suri2021}. This symmetry breaking effectively disappears at higher $Re$, however, as discussed below.

To generate turbulent flows, we computed solutions of equation \eqref{eq:nse} numerically using a pseudo-spectral method. We used a variable-time-step Runge–Kutta–Fehlberg scheme where spatial derivatives and all linear terms were computed in Fourier space and the nonlinear term was computed in physical space. Additionally, we used a two-thirds dealiasing scheme for numerical stability. Most of the results reported here used the grid resolution of $512\times512$. 

\begin{figure}
    \centering
   
       \subfloat[]{\includegraphics[width=0.32\textwidth]{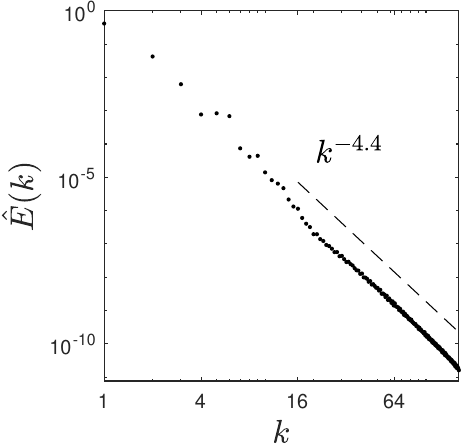}}
       \hspace{1mm}
       \subfloat[]{\includegraphics[width=0.32\textwidth]{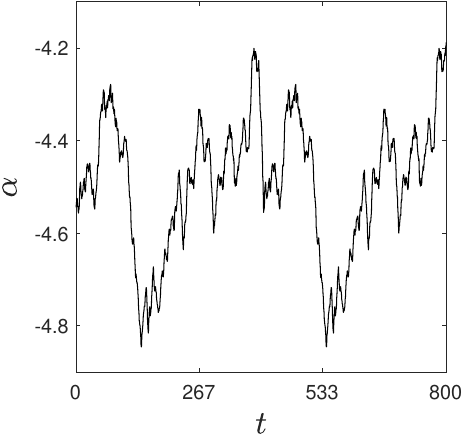}}
       \hspace{1mm}
        \subfloat[]{\includegraphics[width=0.32\textwidth]{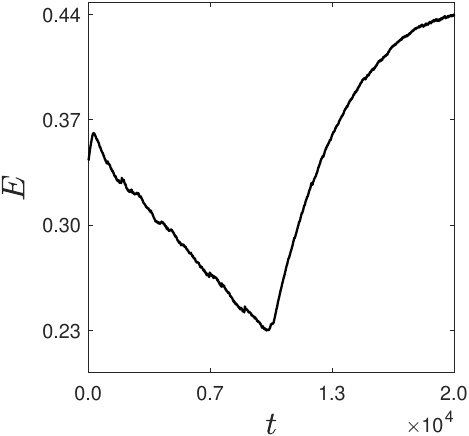}}
    \caption{The energy of turbulent flow. (a) The energy spectrum averaged over a long time interval ($10^3$ nondimensional units) in the asymptotic regime exhibits a clear power-law scaling $E(k)\propto k^\alpha$ (shown as dashed blue line) in the inertial range. (b) The exponent $\alpha$ of the power law computed for energy spectra averaged over a characteristic time scale $T_c$. (c) Variation in the energy over a long time interval. Note that $t=0$ in different plots corresponds to arbitrary and different times.
    }
    \label{f:spectrum}
\end{figure}

In this study we set the Reynolds number to a relatively high value of $Re=10^5$ in order for structure on a broad range of scales to develop, as illustrated by \autoref{f:spectrum}(a). The energy spectrum is found to exhibit a clear power-law scaling $E(k)\propto k^\alpha$ over at least a decade in the wavenumbers ($16\le k\le 170$). This scaling indicates the presence of an inertial range characteristic of fully developed turbulence and clear separation between the $O(1)$ length scale of the forcing and the Taylor microscale $k_t^{-1}$. The exponent $\alpha\approx-4.5$ of the power law is found to differ substantially from the value of $-3$ predicted by the classical theory of turbulent cascades in 2D turbulence developed by \citet{kraichnan1967}, \citet{leith1968}, and \citet{batchelor1969}. This discrepancy is well-known \citep{boffetta2012} and demonstrates the limitations of the KLB theory in properly accounting for the effect of coherent structures on the direct cascade, despite its acknowledgement of the nonlocal nature of interactions \citep{kraichnan1967,kraichnan1971}. Indeed, \autoref{f:spectrum}(b) shows that, although the energy spectrum still retains the power-law scaling when averaged over a short time interval, the exponent $\alpha$ exhibits substantial fluctuations in time reflecting changes in the large-scale structure of the flow.

\begin{figure}
    \centering
       \subfloat[]{\includegraphics[height=0.33\textwidth]{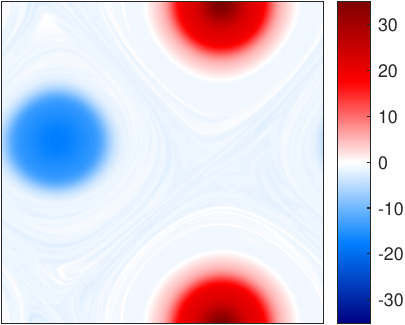}}
       \hspace{5mm}
       \subfloat[]{\includegraphics[height=0.33\textwidth]{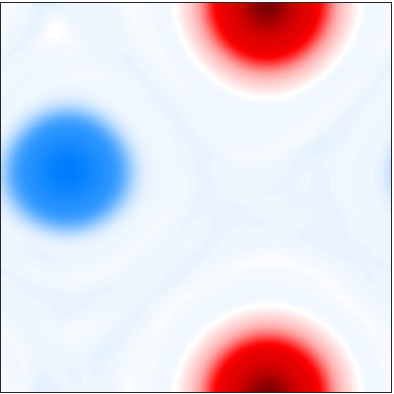}} \\ 
       \subfloat[]{\includegraphics[height=0.33\textwidth]{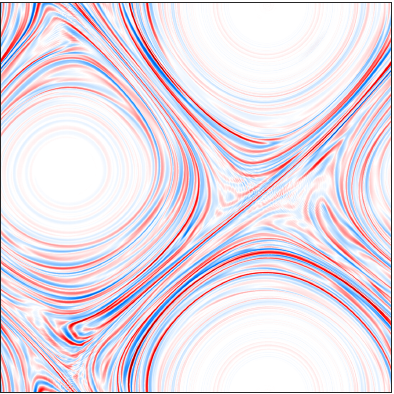}}
       \hspace{16mm}
       \subfloat[]{\includegraphics[height=0.33\textwidth]{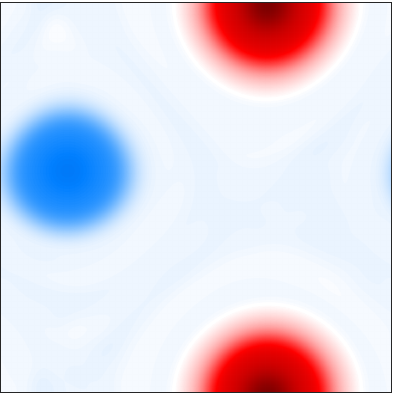}}
    \caption{A typical snapshot of turbulent vorticity field $\omega$.
    The corresponding large-scale flow $\hat{L}_{16}\omega$ (b) and small-scale flow $(1-\hat{L}_{24})\omega$ (c). Panel (d) shows a snapshot of the converged time-periodic ECS. 
    This and all subsequent plots use the same colorbar. For solutions of the Euler equation, such as that shown in panel (d), vorticity scale is arbitrary due to scaling invariance. Hence the color bar shown in panel (a) can be used to interpret vorticity fields shown in all subsequent figures.}
    \label{f:scales}
\end{figure}

The general structure of coherent components of turbulent flow at large and small scales can be easily recognized by applying a Fourier filter. For illustration, we use here a low-pass Fourier filter, denoted by the linear operator $\hat{L}_k$, which corresponds to a smoothed circular window of radius $k$ in Fourier space. We arbitrarily consider wavenumbers $k\le 16$ as representing large scales and $k\ge 24$ as representing small scales. For reference, taking dealiasing into account, our numerical simulations resolve spatial frequencies from the lowest, $k_{\rm min}=k_0$, to the highest, $k_{\rm max}=\lfloor 512/3\rfloor= 170$. The wavenumber associated with the Taylor microscale at which viscous effects become important is somewhat higher, $k_t\sim Re^{1/2}=O(300)$.
As \autoref{f:spectrum}(a) illustrates, we find power-law scaling over the entire wavenumber range corresponding to small scales while, for large scales, significant deviations from a power law are observed. 

A typical snapshot of turbulent flow is shown in panel (a) of \autoref{f:scales}. Panels (b) and (c) show, respectively, the large-scale flow $\hat{L}_{16}\omega$ and the small-scale flow $(1-\hat{L}_{24})\omega$. The large-scale component, for sufficiently-low-frequency forcing such as the one considered here, features a pair of counter-rotating vortices. This is an expected result of the inverse cascade that accumulates the energy in the largest scales accessible to the flow. Small scales represent filaments of vorticity which mainly occupy the space between the two vortices. Stretching (or thinning) of these vorticity filaments in the hyperbolic regions of the large-scale flow is believed to be a key physical mechanism behind the direct cascade \citep{chen2003}.

For higher-frequency forcing, the vortices at the scale closer to that of the forcing become more prominent, and the spectrum has two separate scaling regions, one between the domain scale and the forcing scale controlled by the inverse cascade and another between the forcing scale and the Taylor microscale controlled by the direct cascade. This more complicated situation is outside the scope of the present study. 


The spatial resolution used in our simulations, which may be considered low by modern standards, was motivated by the need to evolve the flow over extremely long time scales.
As illustrated in \autoref{f:spectrum}(c), for the high $Re$ considered here, it takes on the order of $10^5$ nondimensional time units for the flow to come to a statistical equilibrium and for the energy $E=\|{\bf u}\|^2/2$ to approach its long-term average $\overline{E}\approx 0.44$. Here and below, we use the bar to denote the temporal mean and the $L_2$-norm defined by the spatial mean
\eb\label{eq:E}
\norm{{\bf u}}^2=\frac{1}{4\pi^2}\int_0^{2\pi}\int_0^{2\pi}{\bf u}\cdot{\bf u}\,dxdy.
\en
The results presented here describe the statistically stationary turbulence that is found after the long transients have died down {(we initialize the simulation using a small random perturbation about ${\bf u}=0$).} 
A sample movie of turbulent flow in the asymptotic regime is provided as supplementary material. 

To fully resolve the flow at the target $Re$, one formally needs to use an $n\times n$ grid with $n\gtrsim 3k_t\approx 10^3$. Hence, simulations on a $2048\times2048$ grid can be considered fully-resolved. While a number of previous studies of two-dimensional turbulence have used simulations with this resolution \citep{smith1993,bracco2000}, they computed flows over time intervals many orders of magnitude shorter than those considered here. To make sure that our numerical results are representative of fully-resolved simulations of turbulence, we compared them to those obtained on a $2048\times2048$ grid and found the differences to be negligible on the time scale $T_c=10$ characteristic of the recurrent solutions discussed below. We also found the flows, and the recurrence diagrams discussed below, computed on $512\times 512$ and $2048\times2048$ grids, to remain qualitatively similar on time scales of $O(10^3)$. 

\section{$Re\to\infty$ limit}\label{sec:limit}

It is well-known that the viscous term in the Navier-Stokes equation represents a singular perturbation in the limit of $Re\to\infty$. As a result, the solutions of the Euler equation
\eb\label{eq:euler}
\partial_t\omega+ \left({\bf u}\cdot\nabla\right)\omega =0,
\en
which describes the inviscid limit, may have properties that are dramatically different from those of the Navier-Stokes equation. In the inviscid limit, vorticity is materially conserved, giving rise to an infinite number of conserved quantities such as 
\eb
I_n=\frac{1}{4\pi^2}\int_0^{2\pi}\int_0^{2\pi}\omega^n\,dxdy
\en
with integer $n\ge 1$ \citep{eyink1996,clercx2009}, in addition to the conserved energy $E$. Hence, by Noether's theorem, the Euler equation has an infinite number of continuous symmetries and inviscid flows belong to infinite-dimensional solution families spanned by continuous parameters $\sigma_n$, $n=1,2,\cdots$, associated with the corresponding symmetries. On the other hand, the Navier-Stokes equation generally breaks all of these continuous symmetries (except for the temporal translation, so long as the forcing is time-independent) retaining only the discrete translational symmetries of the forcing, if there are any.

As a result of their different symmetries, the Euler equation has a much broader variety of solutions than the Navier-Stokes equation. Consequently, it is natural to ask which solutions of the Euler equation are physically relevant, i.e., correspond to a solution of the Navier-Stokes equation at a high, but finite $Re$. For steady flows with closed streamlines, this limit was originally considered by \citet{batchelor1956} who derived an integral condition on the stream function of the Euler flow. \citet{okamoto1994} has subsequently shown that Batchelor's criterion is equivalent to a solvability condition for this singularly perturbed problem.

This idea can be developed further to obtain several specific, interpretable constraints on the dynamically relevant solutions of the Euler equation, not only steady, but also time-periodic. Let $[{\bf u}_0,p_0]$ be a solution of the Euler equation and  $[{\bf u},p]=[{\bf u}_0,p_0]+Re^{-1}[{\bf u}_1,p_1]$ be a solution of the Navier-Stokes equation \eqref{eq:nseu} with $Re\gg 1$. The perturbation $[{\bf u}_1,p_1]$ satisfies the equation
\begin{align}
    \label{eq:u1} 
    \hat{N}[{\bf u}_1,p_1]=[{\bf H},0],
\end{align}
where we have defined
\begin{align}
    \label{eq:LF} 
    \hat{N}[{\bf u}_1,p_1]&=[\partial_t {\bf u}_1 + ({\bf u}_0\cdot\nabla){\bf u}_1+(u_1\cdot\nabla){\bf u}_0+\nabla p_1,\nabla\cdot{\bf u}_1],\nonumber\\
    {\bf H} &= \nabla^2{\bf u}_0+{\bf h},
\end{align}
and assumed that {\ ${\bf h}= Re\,{\bf f}$} is $O(1)$. 

Subject to the proper boundary conditions in space (and, for time-periodic solutions, in time), equation \eqref{eq:u1} defines a boundary value problem for the perturbation $[{\bf u}_1,p_1]$. Since the linear operator $\hat{N}$ has null eigenvalues associated with every continuous symmetry of the Euler equation, where ${\bf e}_n=[\partial {\bf u}_0/\partial \sigma_n,\partial p_0/\partial \sigma_n]$ are the corresponding eigenfunctions, the boundary value problem is ill-posed and only has solutions provided the solvability condition 
\begin{align}
    \label{eq:solv}
    \langle\partial {\bf u}_0/\partial \sigma_n,{\bf H}\rangle&=0
\end{align}
is satisfied for each of the group parameters $\sigma_n$, where 
\begin{align}
    \langle{\bf u},{\bf v}\rangle=\int_\Omega({\bf u}\cdot{\bf v})d\Omega
\end{align}
and the integral is taken over the entire domain in space (and, for time-periodic solutions, time) on which the inviscid solution $[{\bf u}_0,p_0]$ is defined.

In particular, the scaling symmetry of the Euler equation implies that ${\bf u}_0|_{\sigma_t}=e^{\sigma_t}{\bf u}_0|_{\sigma_t=0}$ and $p_0|_{\sigma_t}=e^{2\sigma_t}p_0|_{\sigma_t=0}$ is a solution for any real $\sigma_t$. Hence, ${\bf e}_t=[{\bf u}_0,2p_0]_{\sigma_t=0}$, and the corresponding solvability condition reduces to
\begin{align}
    \langle{\bf u}_0,{\bf h}\rangle=-\langle{\bf u}_0,\nabla^2{\bf u}_0\rangle.
\end{align}
For $Re\gg 1$, we can replace the inviscid solution with the viscous one yielding, to leading order in $Re^{-1}$,
\begin{align}
    \label{eq:solvt} 
    \langle{\bf u},{\bf f}\rangle=-Re^{-1}\langle{\bf u},\nabla^2{\bf u}\rangle,
\end{align}
so the energy injection by the forcing has to balance the energy dissipation by viscosity at every instant for steady-state solutions and over one period for time-periodic solutions of the Navier-Stokes equation. While energy conservation clearly has to be satisfied for steady and time-periodic viscous flows, our numerical simulations show that, in the asymptotic regime, turbulent flow also satisfies this condition, with the energy becoming essentially time-independent or slowly varying, as Figure \ref{f:spectrum}(c) illustrates. 
Moreover, for smooth solutions of the Euler equation describing large-scale flows, $\langle{\bf u}_0,\nabla^2{\bf u}_0\rangle=O(1)$. Hence, so long as turbulent flow field is close to such an inviscid solution, we have $\langle{\bf u},\nabla^2{\bf u}\rangle\approx \langle{\bf u}_0,\nabla^2{\bf u}_0\rangle$, hence ${\bf u}$ remains essentially orthogonal to the forcing field.

Continuous translational symmetry in the $x$-direction corresponds to the Goldstone mode ${\bf e}_x=[\partial_x{\bf u}_0,\partial_xp_0]$ and yields the solvability condition
\begin{align}
\label{eq:energy_injection}
    \langle\partial_x{\bf u}_0,{\bf h}\rangle=-\langle{\bf u}_0,\partial_x{\bf h}\rangle=-\langle\partial_x{\bf u}_0,\nabla^2{\bf u}_0\rangle.
\end{align}
For a corresponding viscous solution ${\bf u}$, we find, to leading order in $Re^{-1}$, 
\begin{align}
    \label{eq:solvx} 
    \langle{\bf u},\partial_x{\bf f}\rangle=-Re^{-1}\langle\partial_x{\bf u},\nabla^2{\bf u}\rangle.
\end{align}
Again, since $\langle\partial_x{\bf u},\nabla^2{\bf u}\rangle\approx\langle\partial_x{\bf u}_0,\nabla^2{\bf u}_0\rangle =O(1)$ for viscous flows that are close to smooth solutions of the Euler equation, ${\bf u}$ should be (nearly) orthogonal to $\partial_x{\bf f}$. Coupled with the condition \eqref{eq:solvt}, this implies local continuous translational symmetry of the viscous flow field. Similar statements can be made regarding translational symmetry in the $y$-direction. Therefore, although forcing generally breaks continuous translational symmetry at finite $Re$, this symmetry should be effectively restored, at least for infinitesimal shifts, in the limit $Re\to\infty$ {\  for the forcing profiles that do not significantly affect the structure of the large-scale flow}. The same conclusions should apply to a turbulent flow field at high $Re$, so long as it stays in the neighborhood of smooth inviscid solutions.

\begin{figure}
    \centering
\subfloat[]{\includegraphics[width=0.88\textwidth]{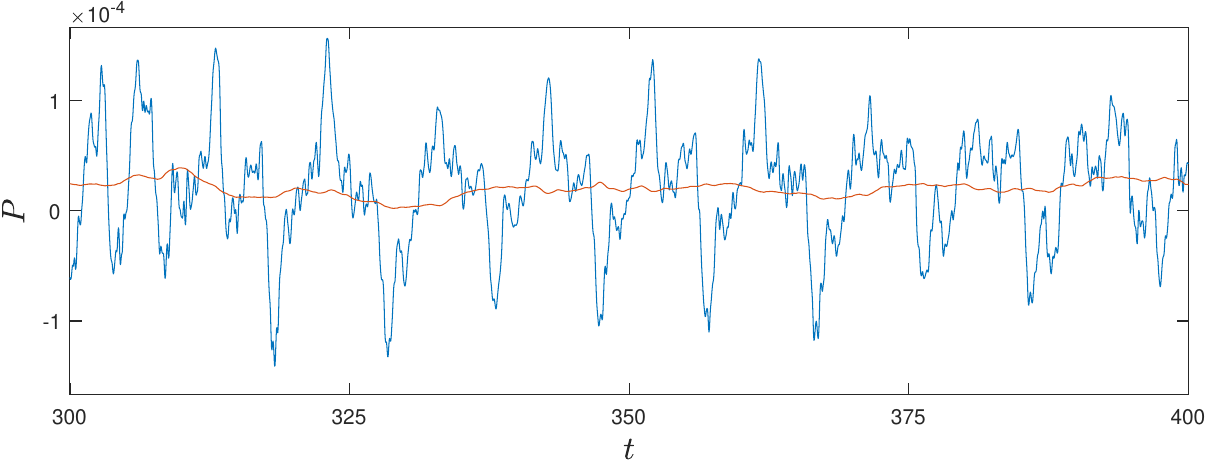}}\\
\subfloat[]{\includegraphics[width=0.44\textwidth]{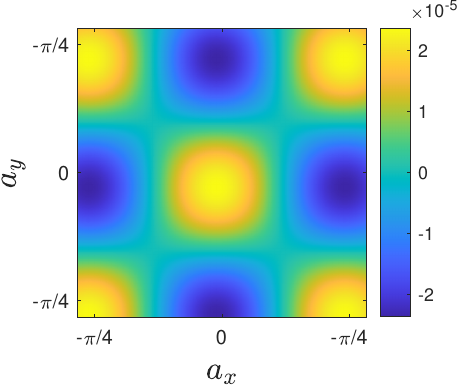}}  
\subfloat[]{\includegraphics[width=0.44\textwidth]{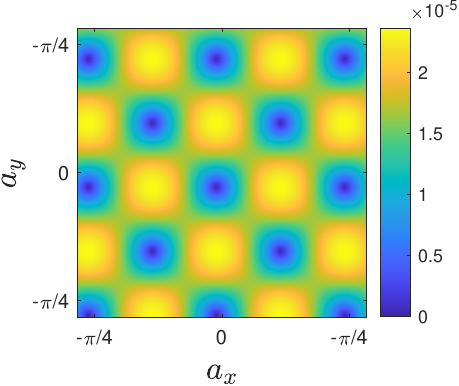}}
    \caption{The energy injection $P$ as a function of time (a). The instantaneous value is shown in blue and the running average computed over the apparent period of the large-scale flow is shown in orange. The temporal average of $P$ (b) and $Q$ (c) computed over the entire time interval $300<t<400$ as a function of the shift ${\bf a}$.}
    \label{f:symm}
\end{figure}

The implications of the solvability conditions for turbulent flow can be quantified by computing the rate of energy injection
\eb\label{eq:Power}
P=\frac{1}{4\pi^2}\int_0^{2\pi}\int_0^{2\pi}{\bf u}\cdot{\bf f}\,dxdy
\en
and the measure of local translational symmetry breaking
\eb\label{eq:translational}
Q=\frac{1}{4\pi^2N}\left|\int_0^{2\pi}\int_0^{2\pi}\sum_iu_i\cdot\nabla f_i\,dxdy\right|.
\en
Figure \ref{f:symm} shows the results for a portion of turbulent trajectory, $300<t<400$, for which the large-scale flow is nearly time-periodic with the apparent period of $T\approx 10$, as indicated by the instantaneous value of $P$ in panel (a). Assuming there is a nearby smooth, time-periodic solution of the Euler equation, we should expect the average of $P$ computed over that period to be $O(Re^{-1})$ for all times, and this is what we find. The temporal average becomes almost constant, with the magnitude representing the balance between the energy injection and viscous dissipation. Moreover, the temporal averages of both $P$ and $Q$ shown in panels (b) and (c) are found to be $O(Re^{-1})$ not only for the turbulent flow ${\bf u}({\bf x},t)$ but also for its shifted version ${\bf u}({\bf x}-{\bf a},t)$, where the shift ${\bf a}$ is allowed to vary continuously over one period of the forcing in both directions, i.e., $-\pi/4<a_x,a_y<\pi/4$. This suggests that, in the limit $Re\to\infty$, the large-scale flow recovers continuous translational symmetry with respect to finite, not just infinitesimal, shifts. 

\section{Numerical results}\label{sec:results}

Temporal evolution of the large-scale flow can be conveniently analyzed by inspecting {a suitably defined recurrence function \citep{cvitanovic2010,lucas2015}}:
\eb\label{eq:recfun}
R(t,\tau) = \min_{\bf a}\norm{\hat L_{16} [{\bf u}({\bf x},t)- 
{\bf u}({\bf x}+{\bf a},t-\tau)]}.
\en
To avoid explicit minimization over the shift ${\bf a}$, effective continuous translational symmetry of the large-scale flow was reduced using Fourier-mode slicing \citep{budanur2015}. Specifically, we shifted ${\bf u}({\bf x},t)$ in both spatial directions so that the phase $\phi_x$ ($\phi_y$) of the first Fourier mode ${\bf k}=(1,0)$ (${\bf k}=(0,1)$) becomes zero before the filtering is applied and the norm is computed in Equation \eqref{eq:recfun}. The components of the shift ${\bf a}$ were then reconstructed from the original phases, e.g., $a_x=\phi_x(t)-\phi_x(t-\tau)$.
Deep (relative to the mean value) minima of $R(t,\tau)$ which correspond to $|{\bf a}|\ll 1$ represent segments of turbulent flow that come close to equilibria or time-periodic states, while deep minima with $|{\bf a}|=O(1)$ correspond to relative equilibria (traveling waves) or relative periodic orbits. 

\begin{figure}
    \centering
    {\includegraphics[height=0.2\textwidth]{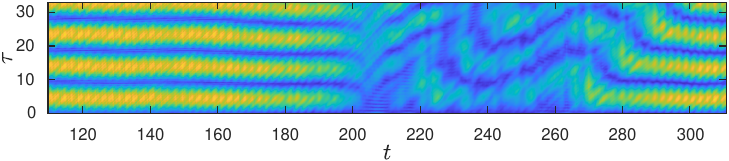}}
    \hspace{1mm}
    \vspace{2mm}\\
    {\includegraphics[height=0.2\textwidth]{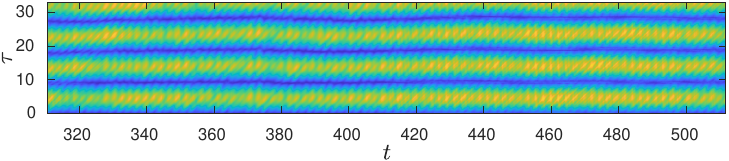}}
    \hspace{1mm}
    \vspace{2mm}\\
    {\includegraphics[height=0.2\textwidth]{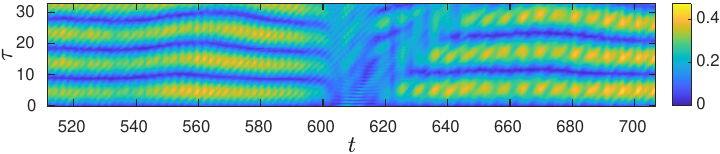}}
    \caption{Recurrence function $R(t,\tau)$ for a Fourier-filtered turbulent flow. The minima at multiples of $\tau\approx 10$ show that large-scale flow exhibits nearly periodic dynamics with a period $T\approx 10$ over extremely long intervals. }
    \label{f:rec1}
\end{figure}

A representative example of such recurrence function is shown in \autoref{f:rec1}. Note that qualitatively similar recurrence functions are produced by DNS on finer meshes (e.g., {$2048\times2048$)}.
Deep minima of $R(t,\tau)$ at integer multiples of $\tau\approx 10$ which, for this time interval, correspond to ${\bf a}\approx 0$, suggest that the large-scale component of the turbulent flow closely follows (or shadows, in the language of dynamical systems) a time-periodic solution of the unforced Euler equation \eqref{eq:euler} with a period $T\approx10$. The corresponding numerically exact solutions can be found using the tuples $\{\bf u({\bf x},t),\tau,{\bf a}\}$ representing the minima as initial conditions for a Newton-GMRES solver accounting for translational symmetry of the Euler equation. A detailed description of such a solver is provided, for instance, by \citet{marcotte2015}. {\  In this study, our focus is entirely on smooth solutions describing large-scale flows. Hence, initial conditions should be prepared by smoothing the flow field ${\bf u}({\bf x},t)$ to eliminate the fine structure, as discussed in the Appendix. Smoothing also helps speed up convergence and improve the success rate of the solver.} For illustration, Figures \ref{f:scales}(b) and \ref{f:scales}(d) show, respectively, snapshots of the vorticity field representing an initial condition {\ (here the large-scale flow without additional smoothing)} and the almost indistinguishable converged solution of the Euler equation.

\begin{figure}
   \centering
    {\includegraphics[height=0.2\textwidth]{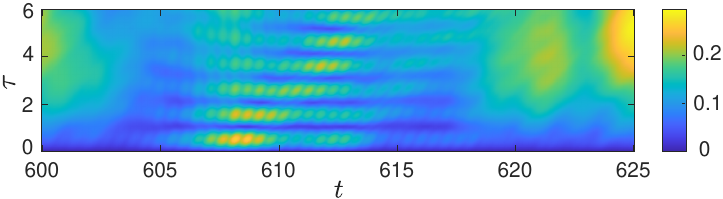}}
    \caption{A zoomed-in version of \autoref{f:rec1} showing a shorter interval of nearly time-periodic dynamics with a period $T\approx 1$.}
    \label{f:rec2}
\end{figure}

Another example of near-recurrences of turbulent flow is shown in \autoref{f:rec2}, which corresponds to the zoomed-in version of \autoref{f:rec1}. In particular, a dark blue triangular region at $t\approx603$ and $\tau\lesssim 2$ represents a short interval when the large-scale flow is nearly stationary. The minima of the recurrence function at $605\lesssim t\lesssim615$ and $\tau\approx 1$ represent a longer interval when the large-scale flow is nearly time-periodic. The corresponding solutions of the Euler equation and their properties are discussed in detail in the subsequent sections. 

\subsection{Equilibria}

Equilibria are the simplest type of recurrent solutions. Perhaps the best-known example is Taylor-Green vortex equilibria which, in two-dimensions, are described by
\begin{align} \label{eq:tgv}
\psi&=A\cos(\gamma(x+a_x))\cos(\gamma(y+a_y)),\nonumber\\
\omega&=-A\gamma^2\cos(\gamma(x+a_x))\cos(\gamma(y+a_y)).\nonumber\\
\end{align} 
Due to the translational symmetry of the Euler equation, the two constant phases $a_x$ and $a_y$ are arbitrary. The constant amplitude $A$ (the analogue of the energy $E$) as well as wavenumber $\gamma$ are also arbitrary due to the scaling symmetry of the Euler equation which, on an unbounded domain, implies that if ${\bf u}({\bf x},t)$ (and $\omega({\bf x},t)$) is a solution of the Euler equation, then so is $\gamma^{-1}\lambda{\bf u}(\gamma {\bf x},\lambda t)$ (and $\gamma^{-1}\lambda^2\omega(\gamma{\bf x},t)$) for any $\lambda$ and $\gamma$. Spatial periodicity restricts $\gamma$ to integer multiples of 1 or $1/\sqrt{2}$, but $\lambda$ can take a continuum of values.

As discussed in Section \ref{sec:limit}, each of the Taylor-Green vortices should belong to a family of equilibria spanned by an infinite number of continuous parameters including $a_x$, $a_y$, and $\lambda$.
Other continuous parameters describe the shape of the vortices. These shape parameters correspond to Lie point symmetries of the Euler equation \citep{Liu2019}. For instance, on an unbounded domain, a single circular vortex
\begin{align} 
\omega=F(r),\qquad r=\sqrt{x^2+y^2}
\end{align} 
is a solution to the steady-state version of the Euler equation
\eb \label{eq:eqcond}
\left({\bf u}\cdot\nabla\right)\omega = 0
\en
for any choice of the shape function $F(r)$.
Indeed, we can expand $F(r)$ in a basis of, say, Bessel functions, with the infinite set of Fourier-Bessel coefficients playing the role of continuous shape parameters.

For our purposes, it is convenient to classify different solution families using topology of their streamlines following \citet{Moffatt1987}. For instance, the Taylor-Green vortex with \begin{align}\label{eq:tgv2}
    \omega=\cos x+\cos y
\end{align}
is obtained by setting $\gamma=1/\sqrt{2}$ and ${\bf a}=0$ in Equation \eqref{eq:tgv} and rotating the coordinate system by $\pi/4$. It belongs to a family of solutions featuring two counter-rotating vortices which also includes an infinite number of other equilibria unrelated by either translational or scaling symmetry. 
\begin{figure}
    \centering
    \subfloat[]{\includegraphics[width=0.3\textwidth]{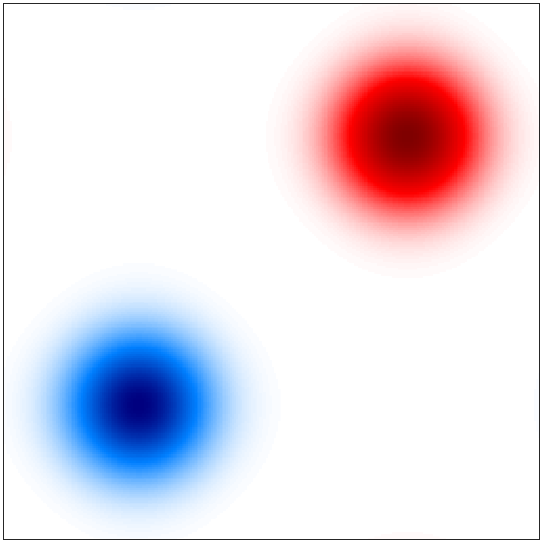}}\hspace{4mm}
    \subfloat[]{\includegraphics[width=0.3\textwidth]{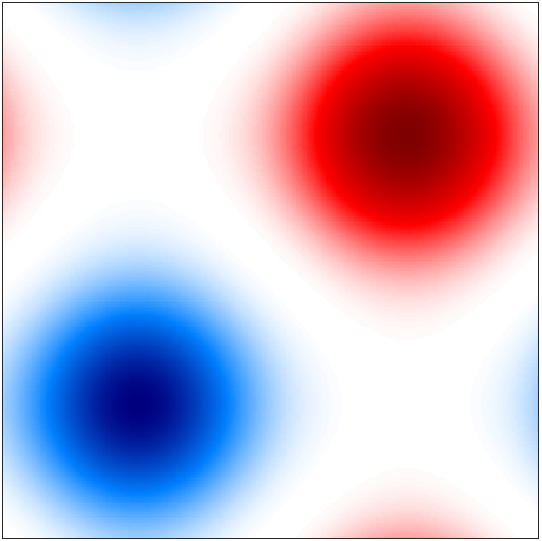}}\hspace{4mm}
    \subfloat[]{\includegraphics[width=0.3\textwidth]{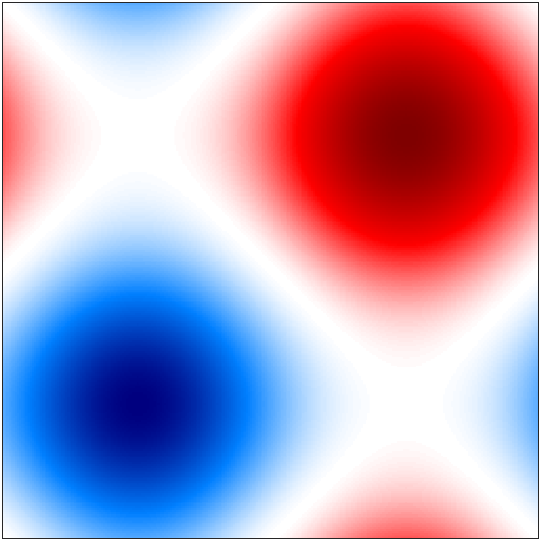}}
    \caption{A family of symmetric equilibria, computed via homotopy between the state shown in (a) and a Taylor-Green vortex \eqref{eq:tgv2} (not shown). Intermediate states are shown in panels (b) and (c).}
    \label{f:eqfam1}
\end{figure}

These solutions can be found either by applying Newton-GMRES solver to flows time-integrated over a short time interval or by solving Equation \eqref{eq:eqcond} directly using GMRES iterations for different initial conditions. For instance, using the state $\omega=(\cos x+\cos y)^4\,\sign(\cos x+\cos y)$ as a non-equilibrium initial condition, we found the equilibrium shown in Figure \ref{f:eqfam1}(a). We then used homotopy {\  (see Appendix)} to construct a continuous family of equilibria connecting that numerical solution with the analytical solution \eqref{eq:tgv2}. {\ Note that homotopy uses known smooth solutions to generate new smooth initial conditions, so no additional smoothing is needed prior to Newton-GRMES iterations.} Representative intermediate equilibria belonging to this family are shown in \autoref{f:eqfam1}(b) and \ref{f:eqfam1}(c). {The members of this family feature two symmetric, counter-rotating vortices and differ mainly in the width of the vortices.} 

\begin{figure}
    \centering
    \subfloat[]{\includegraphics[width=0.3\textwidth]{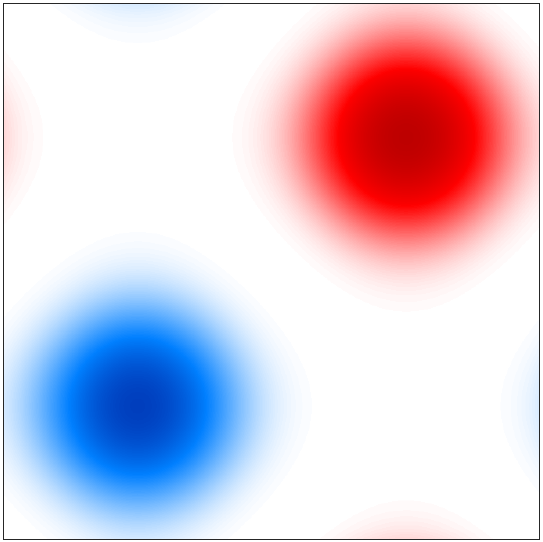}}\hspace{4mm}
    \subfloat[]{\includegraphics[width=0.3\textwidth]{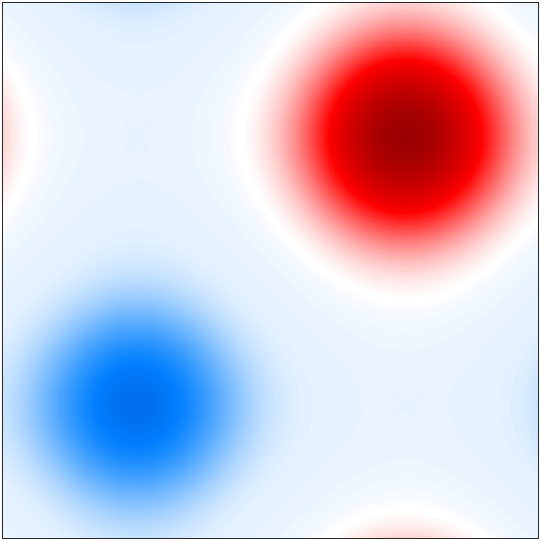}}\hspace{4mm}
    \subfloat[]{\includegraphics[width=0.3\textwidth]{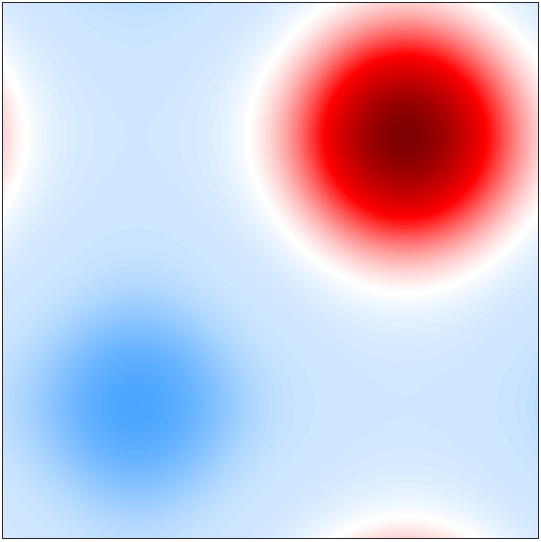}}
    \caption{A family of asymmetric equilibria, computed via homotopy between the states shown in panels (a) and (c). An intermediate state is shown in panel (b).}
    \label{f:eqfam2}
\end{figure}

A different continuous family of equilibria can be obtained by breaking the reflection symmetry between the two vortices. For instance, adding a Gaussian patch of vorticity to the reflection-symmetric equilibrium state shown in \autoref{f:eqfam2}(a) and then subtracting the mean amplifies one of the vortices and weakens and broadens the other. Reconverging the resulting (non-equilibrium) state using Newton-GMRES solver, we found the asymmetric equilibrium state shown in \autoref{f:eqfam2}(c). We then used homotopy to construct another continuous family of solutions, with a representative intermediate equilibrium belonging to this family shown in \autoref{f:eqfam2}(b).

\begin{figure}
    \centering
    \subfloat[]{\includegraphics[height=0.29\textwidth]{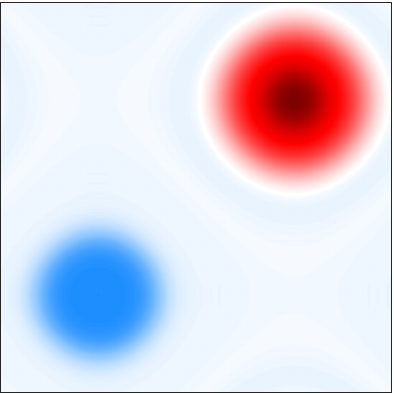}}\hspace{2mm}
    \subfloat[]{\includegraphics[height=0.29\textwidth]{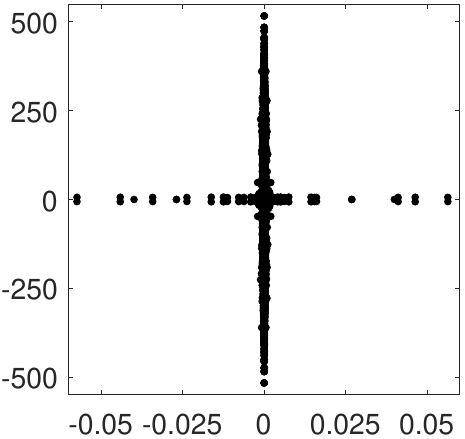}}\hspace{2mm}
    \subfloat[]{\includegraphics[height=0.29\textwidth]{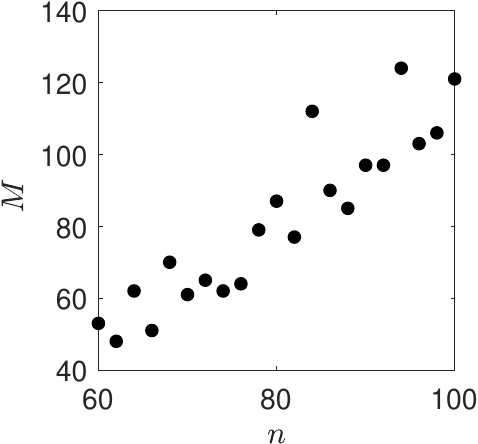}}
    \caption{A dynamically relevant equilibrium computed using a turbulent flow snapshot
    (a) and the corresponding {spectrum of stability eigenvalues $\lambda_i$ (b) computed on a $100\times 100$ grid. The horizontal (vertical) axis corresponds to the real (imaginary) part of $\lambda$. (c) The number $M$ of marginal stability eigenvalues found on an $n\times n$ grid. }}
    \label{f:dyneq}
\end{figure}

Which members of such continuous families, if any, are dynamically relevant for (e.g., are visited by) high-$Re$ turbulent flow needs to be established, however. As discussed previously, such dynamically relevant equilibria can be found using the Newton-GMRES solver seeded with initial conditions identified through recurrence analysis. Good initial conditions correspond to deep minima (compared with the mean value) of $R(t,\tau)$ with a fixed nonvanishing $\tau\ll1$.  One example of such an equilibrium is shown in \autoref{f:dyneq}(a); {\ the corresponding initial condition was obtained by applying spectral and stream function smoothing to the snapshot of turbulent flow at $t\approx 603$ corresponding to Figure \ref{f:rec1}}. This solution is qualitatively quite similar to the equilibria shown in Figures \ref{f:eqfam1} and \ref{f:eqfam2} which also speaks in favor of all these equilibria belonging to one family parameterized by multiple continuous parameters.

The number of continuous parameters for any family will be reflected in the stability spectrum of any member of the family. Since generators of the Lie group describing different continuous parameters commute with the evolution operator, each continuous parameter would correspond to a marginal eigenvalue associated with a corresponding marginal mode. Consider, for instance, the dynamically relevant equilibrium shown in \autoref{f:dyneq}(a). Its stability spectrum computed {in Fourier space using the matlab function eig() is shown in \autoref{f:dyneq}(b). Due to memory constraints, this computation has been performed  on grids up to $100\times100$. Since the total number and accuracy of the eigenvalues depends on both the spatial resolution of the flow field and the accuracy of eig(), it is difficult to estimate the number of continuous parameters precisely.
Let us define, rather arbitrarily, an eigenvalue $\lambda_i$ to be marginal so long as $|\lambda_i|<10^{-5}\ll \max_i(Re(\lambda_i))\approx 0.06$. The number $M$ of such marginal eigenvalues, computed on an $n\times n$ grid, grows roughly linearly with $n$, as illustrated by \autoref{f:dyneq}(c), suggesting that $M\to\infty$ as $n\to\infty$. Note that the limited accuracy of eig() is also responsible for the deviations from the expected inversion symmetry $\lambda\to-\lambda$ in \autoref{f:dyneq}(b) which reflects the time-reversal symmetry of the Euler equation.

The stability spectrum also shows that the equilibrium is only moderately unstable. Unstable perturbations grow on a time scale of $\max(Re(\lambda))^{-1}\approx 17$ comparable to the characteristic time scale {$T_c\approx 10$} associated with turbulent flow. At first sight, this result appears very surprising. Indeed, this equilibrium describes inviscid flow and therefore corresponds to the limit $Re\to\infty$, where ECSs are expected to be very strongly unstable. While that may indeed be the case for recurrent solutions describing small-scale flows (i.e., coherent {\it substructures}), the presence of the inverse cascade in 2D implies that recurrent solutions describing large-scale flows (i.e., coherent {\it structures}) can only be unstable with respect to perturbations with low spatial frequencies. These low-frequency modes are largely aligned with the infinite-dimensional subspace spanned by the marginal modes for {\it any} large-scale Euler equilibrium. This near-alignment is responsible for the relatively small real parts of the eigenvalues associated with  unstable eigenmodes. 

\begin{figure}
    \centering
       \subfloat[]{\includegraphics[width=0.33\textwidth]{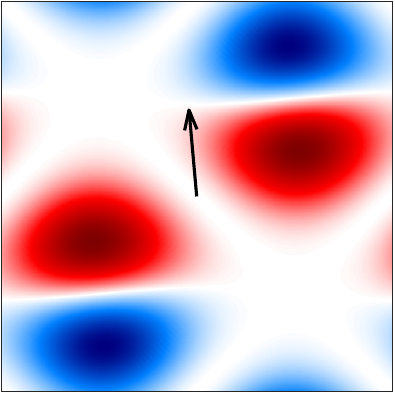}}
       \hspace{10mm}
       \subfloat[]{\includegraphics[width=0.33\textwidth]{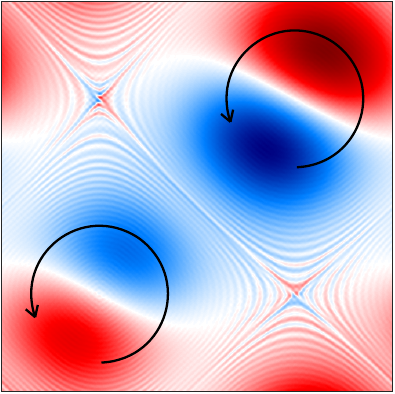}}
       \caption{The leading mode of instability for equilibria calculated using power iteration. (a) The leading eigenfunction for the equilibrium shown in \autoref{f:eqfam2}(a) represents an infinitesimal shift associated with translation of the flow pattern in the direction shown by the arrow. (b) The leading Floquet mode for the equilibrium shown in \autoref{f:eqfam2}(b) represents an infinitesimal shift of the vortex cores associated with circular motion of the flow pattern indicated by the arrows. The snapshot shown corresponds to a particular phase of the oscillation.  }
       \label{f:eqpert}
\end{figure}

Overall, we find that although the large-scale turbulent flow does occasionally visit the neighborhoods of equilibrium solutions of Euler, recurrence analysis shows that these instances are quite rare. In order to understand why that is the case, we computed the leading eigenmodes associated with a pair of representative equilibrium states. 
The most unstable eigenmode corresponding to the symmetric equilibrium from \autoref{f:eqfam2}(a) is shown in \autoref{f:eqpert}(a). It is almost indistinguishable from the marginal mode associated with spatial translation in the direction indicated by the black arrow. Indeed, numerically evolving the equilibrium, we find that it transitions to a traveling wave moving in the corresponding direction. The similarity of the leading eigenmode to the translational marginal mode implies that the relative position of the vortices for the traveling wave does not change in time and hence neither does the flow generated by the vortex array in the co-moving reference frame. As a result, the direction in which the vortices travel never changes either. 

For the asymmetric equilibrium from \autoref{f:eqfam2}(b), the situation is quite different. The leading eigenvalues in this case form a complex conjugate pair, so the the corresponding eigenmode is time-periodic; its snapshot is shown in \autoref{f:eqpert}(b). The spatial structure of this mode is quite interesting and sheds some light on the dynamics of both large- and small-scale flow structures in turbulence. In the regions associated with the vortex cores, the eigenfunction is similar to the translational marginal mode, and represents a shift of the vortices. As time progresses, the corresponding dipolar structures rotate in the direction indicated by the arrows, which suggests that this equilibrium evolves into a time-periodic flow, with the vortex centers executing a circular motion, as confirmed by numerical simulation.

The structure of the eigenmode in the hyperbolic regions does not correspond to a rigid spatial shift, as can be seen by comparison with \autoref{f:eqpert}(a), suggesting that the flow pattern not only translates but also deforms. Instead, the eigenmode features thin vorticity filaments aligned with the streamlines of the base flow. These vorticity filaments are qualitatively similar to those found in turbulent flow (cf. \autoref{f:scales}(c)). This observation suggests that it is the time-dependence of the large-scale flow that is responsible for generation of small scales and the direct cascade overall.

Since symmetric equilibria are rare compared with asymmetric ones (they represent a set of measure zero in the corresponding continuous parameter space), we should expect equilibria to transition predominantly to time-periodic flows. Indeed, the recurrence function (cf. \autoref{f:rec1}) illustrates that the large-scale flow is nearly time-periodic over a significant fraction of the time, but only rarely resembles traveling waves. Both types of solutions are discussed in more detail below. 


\subsection{Traveling waves}

While traveling waves appear to represent time-periodic flows on a domain with periodic boundary conditions, they become equilibria in a co-moving reference frame, i.e., they correspond to relative equilibria. Absolute equilibria discussed in the previous section correspond to the special case of the co-moving frame having a zero velocity. Since the velocity of a generic co-moving frame is quite unlikely to vanish, equilibria are expected to be far less common than traveling waves, which explains why time-independent large-scale flows are so rarely observed in the DNS of turbulence.

\begin{figure}
    \centering
         \subfloat[]{\includegraphics[height=0.33\textwidth]{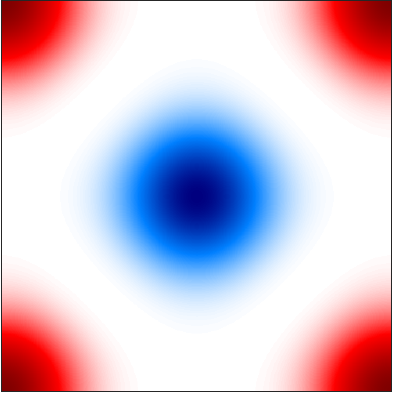}}
       \hspace{10mm}
       \subfloat[]{\includegraphics[height=0.33\textwidth]{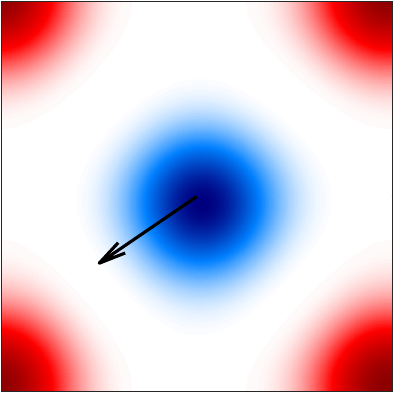}}\\
       \subfloat[]{\includegraphics[height=0.33\textwidth]{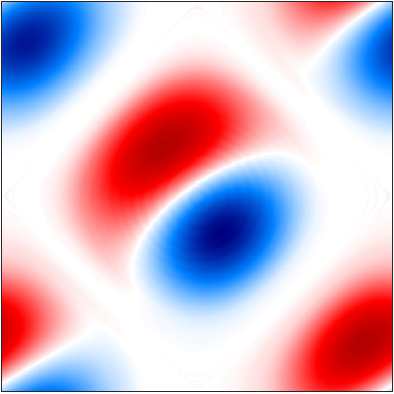}}
       \hspace{8mm}
       \subfloat[]{\includegraphics[height=0.33\textwidth]{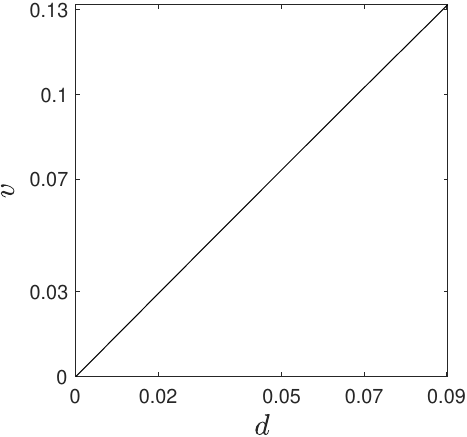}}
    \caption{A continuous family of solutions connecting an equilibrium (a) and a traveling wave solution (b). The marginal mode representing this solution family (c). The speed of the co-moving frame as a function of the state space distance to the equilibrium (d). }
    \label{f:TwtoEq}
\end{figure}

Traveling waves and equilibria of the Navier-Stokes equation are distinct and the velocity of the co-moving frame for traveling waves can only take discrete values \citep{chandler2013,lucas2014,lucas2015}. In contrast, traveling waves and equilibria of the Euler equation are connected, and the velocity of the co-moving frame can take a continuum of values. This is illustrated in \autoref{f:TwtoEq}, which describes a continuous family of traveling wave solutions obtained with the help of homotopy between an equilibrium shown in panel (a){\ , which belongs to the family shown in \autoref{f:eqfam1},} and a traveling wave shown in panel (b). {\  The initial condition for the travelling wave was generated by shifting one of the vortices. Specifically, the vorticity field was updated according to $\omega\to\omega+\varepsilon{\bf n}\cdot\nabla\omega$ in the regions with $\omega>0$, where $\varepsilon\ll2\pi$ and ${\bf n}=(1,-1)$, followed by spectral smoothing, and then integrating the flow in time until visible transients disappeared.} While the corresponding vorticity fields for these two solutions appear extremely similar, they are slightly different. Their difference is described by the marginal mode associated with this continuous family which is shown in panel (c) and represents a relative displacement of the two vortices. Just like arrays of point vortices, equilibrium arrangements of extended vortices require high symmetry balancing advection flows generated by actual or virtual neighbors. When the symmetry of a vortex array is broken, advection causes an overall drift in the pattern. As shown in panel (d), the drift velocity $v$ increases with the degree of asymmetry quantified by the normalized state space distance
\begin{align}\label{eq:disteq}
    d({\bf u})=\frac{\|{\bf u}-{\bf u}_0\|}{\norm{\bf u_0}}
\end{align}
from the equilibrium ${\bf u}_0$.

In the example considered here, the two vortices have the same shape and strength and drift with the same velocity, yielding a traveling wave. When the shapes and/or strengths of the two vortices are different, their drift speeds will not be the same, yielding a gradual change in the relative position of the vortices and therefore a change in the direction of the drift. Hence asymmetric vortices will generally have dynamics that are more complicated than traveling waves; one example is time-periodic states discussed in the next section. We can again invoke the argument that symmetric vortex arrangements are rare, so ECSs in the form of traveling waves should also be relatively uncommon in turbulence. This prediction is indeed born out by recurrence analysis.

We will conclude our discussion of traveling waves by noting that they, just like absolute equilibria, are expected to belong to families with an infinite number of continuous parameters, some of which correspond to translational and scaling symmetry while the rest describe the shape and relative position of the vortices. The nontrivial example shown in \autoref{f:TwtoEq} represents continuous variation in the relative position for vortices with a particular shape. However, given the arbitrary choice of the vortex shape in this example, vortices with any other shape can also be continued to corresponding traveling waves by varying their relative position.

\subsection{Periodic orbits}

As illustrated by the recurrence function shown in \autoref{f:rec1}, turbulent flow exhibits long intervals during which large-scale flow is almost time-periodic in some reference frame. The shift ${\bf a}$ during these intervals tends to be quite small, suggesting that the co-moving reference frame is stationary and, therefore, the large-scale flow is well-described by an unstable periodic orbit solution (UPO) of Euler. The recurrence function shows that there are two distinct types of UPOs characterized by substantially different periods: $T\approx 10$ or $T\approx 1$. A representative example of UPO with period 10.017 is shown in \autoref{f:scales}(d), and a corresponding movie is provided as supplementary material.
This type of UPO features a pair of axially symmetric vortices, just like the equilibria and traveling waves discussed previously. The shapes of the vortices do not change noticeably over the period, while their centers undergo a nearly circular motion. 

A representative example of UPO with the shorter period is shown in \autoref{f:tripole1}(b); a corresponding movie is provided as supplementary material. 
{\ The initial condition was obtained by applying a combination of hyperviscous smoothing and spectral smoothing to a snapshot of a turbulent flow.}
This type of UPO features a pair of vortices only one of which is axially symmetric. The other vortex has a strongly asymmetric shape, e.g., elliptical or, as in the present example, tripolar. We find that the centers of both vortices are essentially stationary, so the time-dependence is associated entirely with the rigid rotation of the asymmetric vortex.
Coherent structures featuring tripolor vortices in particular have been frequently observed in both numerical simulations of 2D turbulence \citep{legras1988} and experiments in rotating tanks \citep{Heijst1991}. Our results show that such coherent structures correspond to numerically exact solutions of the Euler equation.

\begin{figure}
    \centering
    \subfloat[]{\includegraphics[width=0.3\textwidth]{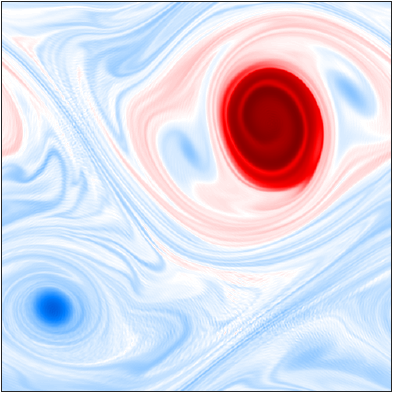}}\hspace{4mm}
    \subfloat[]{\includegraphics[width=0.3\textwidth]{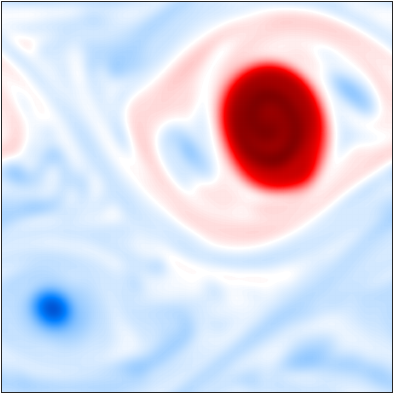}}\hspace{4mm}
    \subfloat[]{\includegraphics[width=0.3\textwidth]{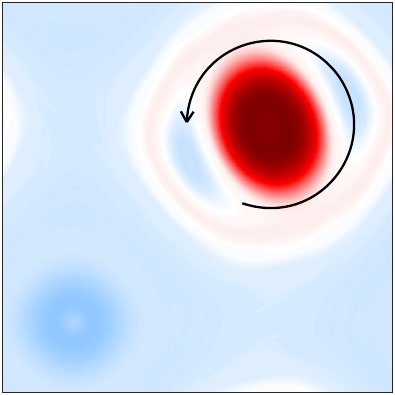}}\\
    \caption{Snapshot of turbulent flow $\omega$ (a), the corresponding large-scale flow $\hat{L}_{16}\omega$ (b) and the converged UPO with the period  $T\approx 1.05$ (c). The arrow indicates the direction of rotation for the tripolar vortex. }
    \label{f:tripole1}
\end{figure}

This is illustrated by \autoref{f:tripole1} which compares a snapshot of DNS of 2D turbulence featuring a tripolar vortex with the corresponding large-scale flow and numerically converged UPO of Euler. It is important to note that {\ smoothing affects the smaller-scale structures of the flow, and}
Newton-GMRES solver is not guaranteed to converge to the ECS that is the closest to the initial condition, so some difference in the vorticity fields shown in panels (b) and (c) is expected.
It is worth noting that solutions featuring a single tripolar vortex on an unbounded spatial domain correspond to relative equilibria of Euler, e.g., equilibria in a co-rotating reference frame. In the present case, time-periodicity arises from breaking of the rotational symmetry of Euler by both the boundary conditions and the presence of another (stationary and axially symmetric) vortex.

Just like equilibrium and traveling wave solutions of the Euler equation, UPOs are found to come in families parameterized by numerous continuous parameters, some of which correspond to translational and scaling symmetry of Euler while others describe the shape and relative position of the vortices. {\ \autoref{f:pofam1} shows one such family of UPOs; a corresponding movie is provided as supplementary material. A reference time-periodic solution shown in \autoref{f:pofam1}(b) was obtained by further smoothing the flow shown in \autoref{f:scales}(b). The solution family was constructed by converging initial conditions obtained by modifying the vorticity field of the reference solution according to $\omega\to|\omega|^\kappa\sign(\omega)$ over a continuous range of parameter $\kappa$. Solutions corresponding to $\kappa=0.5$ and $\kappa=1.8$ are shown in \autoref{f:pofam1}(a) and (c), respectively. This procedure is an alternative to the homotopy that allows constructing a continuous family of solutions using one reference state rather than two.}
The continuous nature of this solution family is seen in \autoref{f:pofam1}(d) which shows the period $T$ and the amplitude
\eb
A = \frac{1}{T}\int_0^T\sqrt{\left(\phi_x(t)-\overline{\phi_x}\right)^2+\left(\phi_y(t)-\overline{\phi_y}\right)^2}\,dt,
\en
of the UPO, which quantifies the degree of unsteadiness of the flow, as a function of the enstrophy $H=I_2$.

\begin{figure}
    \centering
       \subfloat[]{\includegraphics[height=0.33\textwidth]{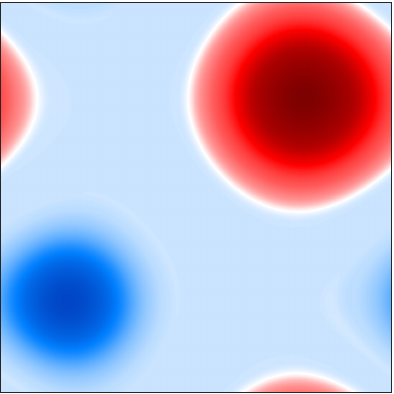}}
      \hspace{12mm}
       \subfloat[]{\includegraphics[height=0.33\textwidth]{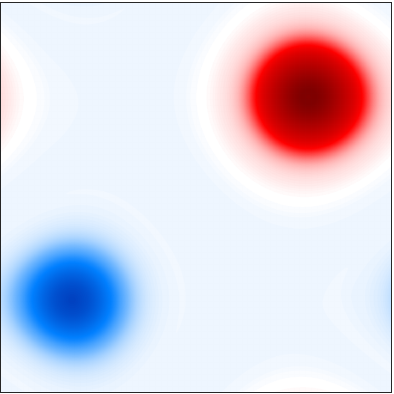}}\\
       \subfloat[]{\includegraphics[height=0.33\textwidth]{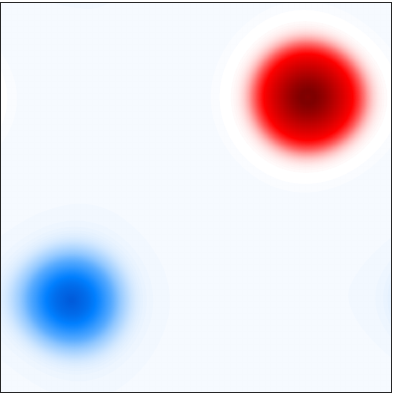}}
      \hspace{4mm}
       \subfloat[]{\includegraphics[height=0.33\textwidth]{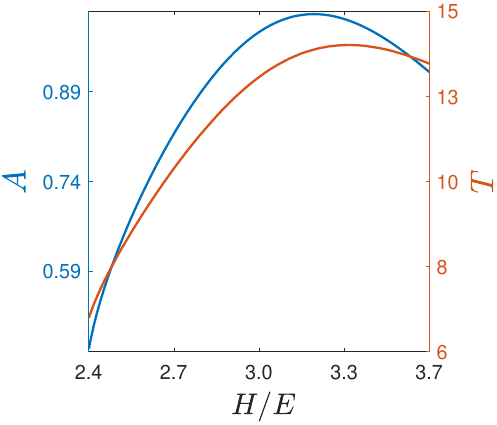}}\\
    \caption{A continuous family of UPOs which reflects variation in the shape of the vortices. Panels (a-c) shows snapshots of representative states from the same family. (d) Dependence of the amplitude and period of UPOs on the normalized enstrophy for this solution family.
    }
    \label{f:pofam1}
\end{figure}

Just like traveling wave solutions, for the Euler equation, UPOs are not isolated from the other types of solutions. \autoref{f:POtoEq} illustrates this for a continuous family of solutions connecting an equilibrium shown in panel (a) and a UPO shown in panel (b); a corresponding movie is provided as supplementary material. {\  The equilibrium featuring two asymmetric vortices was obtained through the same approach as that used to generate the state shown in \autoref{f:eqfam2}(c) and the UPO was obtained from the equilibrium through the same procedure as that used to generate the state shown Figure \ref{f:TwtoEq}(c).} 
The marginal mode describing this family is shown in panel (c) and represents a relative shift of the vortices. The equilibrium here, unlike that shown in \autoref{f:TwtoEq}, features vortices with different shapes. As discussed previously, the shape asymmetry leads to the drift of the vortices in a direction that changes with time. The vortex pattern moves in a circular path, as indicated by an arrow in panel (b), repeating after period $T$. 

\begin{figure}
    \centering
        \subfloat[]{\includegraphics[height=0.33\textwidth]{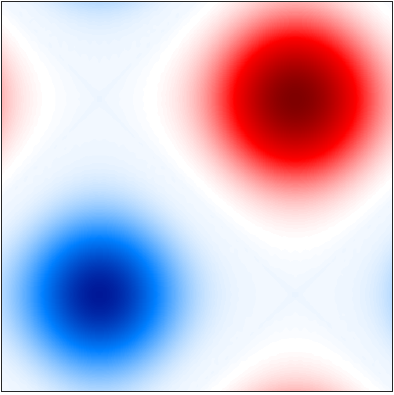}}
       \hspace{12mm}
       \subfloat[]{\includegraphics[height=0.33\textwidth]{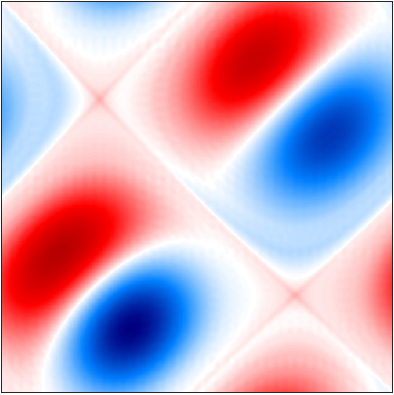}}\\
       \subfloat[]{\includegraphics[height=0.33\textwidth]{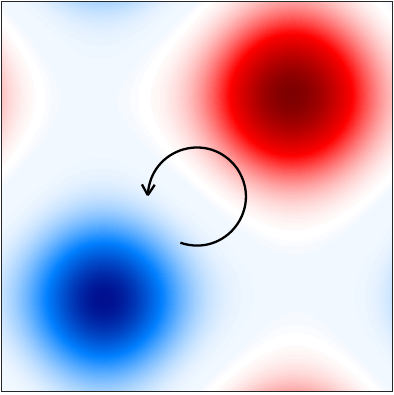}}
       \hspace{5mm}
       \subfloat[]{\includegraphics[height=0.33\textwidth]{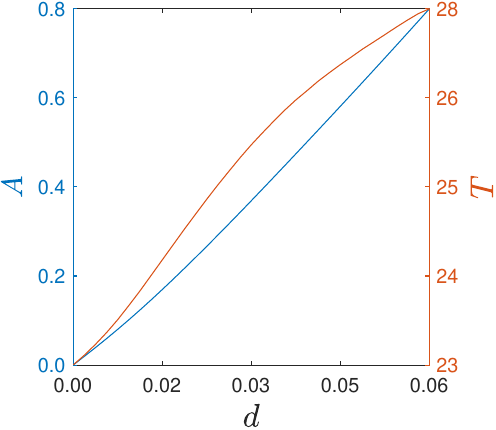}}
    \caption{A continuous family of UPOs connecting the equilibrium shown in panel (a) and a UPO shown in panel (b). The arrow represents the direction in which the pattern moves.  (c) The marginal mode associated with this solution family. (d) The period and amplitude of UPOs as a function of the state space distance to the equilibrium. 
}
    \label{f:POtoEq}
\end{figure}

All UPOs characterized by the overall drift of the pattern that we found are formally pre-periodic: the state at time $t=T/4$ is related to the state at time $t=0$ by a $\pi/2$ rotation about some point in the domain. As a result, the pattern moves in an almost circular trajectory (an example is shown in \autoref{f:path1} in red). The amplitude $A$ effectively describes the radius of the circle.
For the family of UPOs shown in \autoref{f:POtoEq}, the amplitude is shown in panel (d) and is found to vanish at the corresponding equilibrium and grow monotonically as the distance $d$ from that equilibrium increases. As the supplementary movies show, both vortices move along nearly circular trajectories with somewhat different radii, but the vortex pattern remains qualitatively the same at all times.

To illustrate that families of UPOs characterized by such circular motion do indeed describe the dynamics of turbulent flow, we have plotted the position of the vortex pattern describing the large-scale flow, as characterized by the first-Fourier-mode phases $\phi_x$ and $\phi_y$, in \autoref{f:path1}. The trajectory of the turbulent flow in this low-dimensional projection of the state space is nearly time-periodic over the time intervals $77\lesssim t\lesssim 200$ and $270\lesssim t\lesssim 400$. These intervals are separated by an interval $200\lesssim t\lesssim 270$ during which the dynamics are aperiodic. Indeed, during the time-periodic intervals, the pattern moves in a (nearly) circular trajectory (shown as a solid black line) which closely resembles the trajectory of particular UPOs (shown as red circles). During the aperiodic interval, the trajectory (shown as a dashed black line) becomes much more complicated and is not described by any UPO. The observation that the phases $\phi_x$ and $\phi_y$ vary continuously in time instead of taking discrete values is consistent with the analysis of Section \ref{sec:limit} which suggests that, at high $Re$, turbulent flow recovers continuous translational symmetry.

\begin{figure}
    \centering
       {\includegraphics[width=0.5\textwidth]{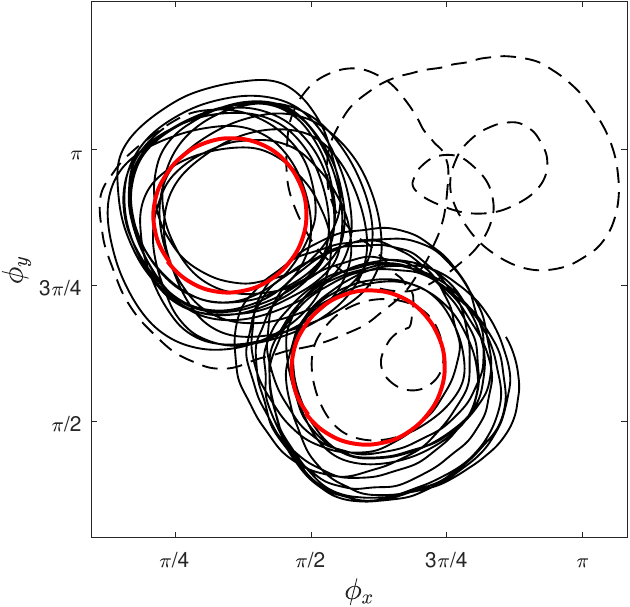}}
    \caption{The position of the vortex pattern describing turbulent large-scale flow (in black) corresponding to \autoref{f:rec1} over the time interval $77\lesssim t\lesssim 400$. Sub-intervals of recurrent (non-recurrent) dynamics are shown as solid (dashed) line. 
    The trajectory of a representative UPO, computed from an initial condition along the recurrent portion of the flow, is shown in red.}
    \label{f:path1}
\end{figure}

As the recurrence function shown in \autoref{f:rec1} further illustrates, the temporal period of the large-scale flow is not constant but varies slowly, and so does the amplitude $A$. Therefore, turbulent flow follows different UPOs. This variation represents a slow drift in the infinite-dimensional parameter space describing a family of the UPOs. Due to the scaling invariance of the Euler equation, one of these continuous parameters is the energy $E$; the energy of dynamically relevant UPOs of the Euler equation should be equal to the energy of the large-scale turbulent flow. Recall that the Euler equation conserves energy, so $E$ is constant for any solution, including UPOs. Since the period of any UPO scales with its energy, $T\propto E^{-1/2}$, a slow variation in $E$ should cause a slow variation in the temporal period. However, the energy is not the only parameter that varies and affects the period. 

In particular, consider the variation in the enstrophy $H$ of the turbulent flow shown as a continuous blue line in \autoref{f:TrajectoryLSECS}, computed after the low-pass filter has been applied, during the time interval where the dynamics are nearly time-periodic. We used appropriately smoothed snapshots of the turbulent flow as initial conditions for the Newton-GMRES solver to compute the corresponding UPOs. The enstrophy describing these UPOs (shown as black circles) is found to track that of the turbulent flow, providing further evidence that the evolution of large-scale flow at high but finite $Re$ can be well-described as a slow drift in the infinite-dimensional space of parameters describing continuous families of unstable time-periodic solutions of Euler.

\begin{figure}
    \centering
    {\includegraphics[width=0.6\textwidth,]{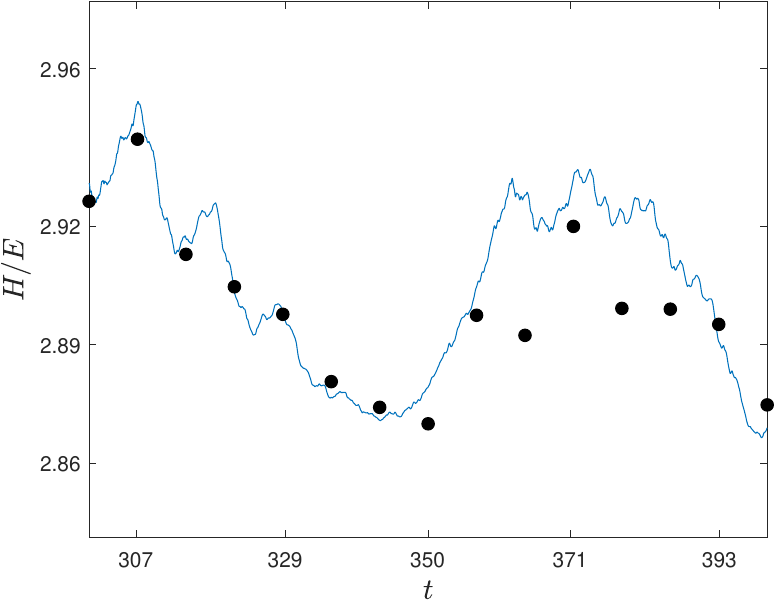}}
    \caption{The normalized enstrophy $H$ of turbulent large-scale flow (in blue) corresponding to the interval $t\in[300,400]$ in Figure \ref{f:rec1}. The symbols correspond to UPOs converged from turbulent flow at the corresponding points in time. }
    \label{f:TrajectoryLSECS}
\end{figure}

In conclusion of this section, let us address the issue of stability of dynamically relevant UPOs which explains why large-scale flow can remain nearly time-periodic over time intervals far exceeding the characteristic time scale $T_c$. Computing stability spectra of UPOs of the Euler equation with meaningful precision is extremely challenging due to the very large number of marginal (unit) Floquet multipliers associated with various continuous parameters and the absence of a spectral gap with unstable multipliers. As a result, standard approaches such as Arnoldi iterations become intractable. To assess the stability of a representative UPO ${\bf u}({\bf x},t)$ with period $T=12.23$, (whose snapshot is shown in \autoref{f:poPert}(a)), we computed the evolution of a random perturbation by time-integrating the Euler equation linearized about this UPO. The magnitude of an infinitesimal perturbation $\delta{\bf u}$ at integral multiples of the period is shown in \autoref{f:poPert}(b). The growth is found to be algebraic rather than exponential over 30+ periods, with the growth rate steadily decreasing over time. This suggests that the leading Floquet multiplier is extremely close to unity and we simply observe transient growth associated with the nonnormality of the evolution operator. Additional validation of this conclusion is provided by the spatial profile of the perturbation which, at long times, becomes virtually indistinguishable from that of the marginal mode $\partial_t{\bf u}$ associated with time translation (cf. \autoref{f:poPert}(c) and (d)). The corresponding perturbation essentially represents a slow drift in the period of the UPO.

\begin{figure}
    \centering
    \subfloat[]{\includegraphics[width=0.33\textwidth]{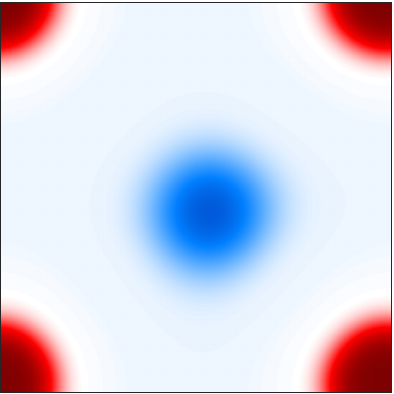}}\hspace{10mm}
    \subfloat[]{\includegraphics[width=0.33\textwidth]{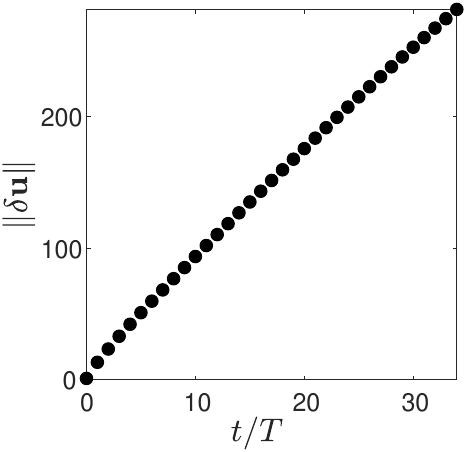}}\\
    \subfloat[]{\includegraphics[width=0.33\textwidth]{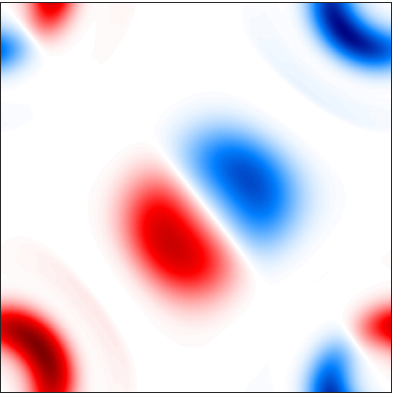}}\hspace{10mm}
    \subfloat[]{\includegraphics[width=0.33\textwidth]{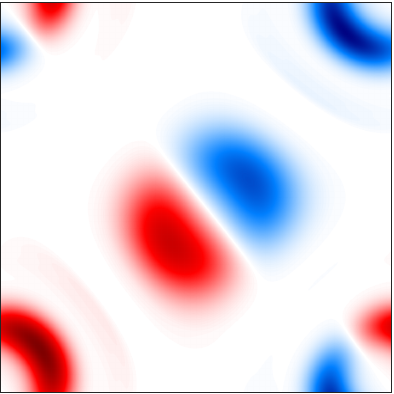}}
    \caption{Stability of a dynamically relevant periodic orbit shown in (a). Transient growth in the amplitude of a random infinitesimal perturbation (b). The shape of the perturbation at long times (c) is nearly identical to the marginal mode $\partial_t {\bf u}$ (d).}
    \label{f:poPert}
\end{figure}

\section{Discussion}\label{sec:disc}

We found that sustained high-$Re$ turbulence in a spatially periodic two-dimensional domain driven by stationary forcing, for extended intervals of time, exhibits nearly time-periodic dynamics on large scales. This observation has important implications for our understanding of the direct cascade. During these intervals, the large-scale flow is well-described by particular weakly unstable time-periodic solutions of the Euler equation. These UPOs all feature a pair of counter-rotating vortices separated by hyperbolic regions of extensional flow which contain most, if not all, of the small-scale structure. The stretching and folding of thin vorticity filaments due to, respectively, the advection by the large-scale flow and its time-dependent nature are a familiar mechanism which generates small-scale structure in chaotic advection of passive tracers by time-periodic flows \citep{cartwright1996,grigoriev2005}. Indeed, the analogy between small-scale vorticity and passive tracers whose evolution is described by formally the same equation, has been noted a long time ago by \citet{kraichnan1975}. However, the realization that flow at a single (e.g., large) scale could be solely responsible for generating a fractal structure characterized by power-law scaling with a non-integral exponent is much more recent \citep{ott1999}. 

The presence of large-scale coherent structures is well-known \citep{clercx2009,boffetta2012} to lead to substantial deviations from the predictions of KLB theory which assumes implicitly or explicitly \citep{kraichnan1975} that small scales add up incoherently. Indeed the presence of large-scale structure implies strong coherence at small scales as well. As our results illustrate, the relative position of the vortices and hence the orientation of the extensional flow remains nearly constant, implying strong correlations between the orientation of vorticity filaments at all scales. These strong correlations explain why the observed power law exponent $\alpha$ deviates so strongly from the KLB prediction for the turbulent flow considered here.

{Rather counter-intuitively, while the properties of the large-scale flow, which control the dynamics of small-scale vorticity and hence the direct cascade, tend to change slowly and smoothly, the scaling exponent changes rather abruptly and unpredictably, as illustrated by \autoref{f:spectrum}(b). This observation echoes a similar one made by \citet{ottino1989} who noticed that transport properties of time-periodic two-dimensional flows can change in a rather irregular fashion when their parameters are varied smoothly. Transport and mixing of passive scalars is a result of Lagrangian chaos, and chaotic sets are known to exhibit irregular dependencies on parameters. Perhaps the best known example of such irregular dependence is provided by the windows of periodic dynamics of the logistic map (characterized by poor mixing) interspersed within the chaotic parameter interval (characterized by good mixing).}

\section{Conclusions}\label{sec:conc}

The present study identified several classes of numerically exact unstable recurrent solutions of the Euler equation on a 2D square domain with periodic boundary conditions: equilibria, traveling waves and periodic orbits. All three types of solutions feature a pair of extended counter-rotating vortices. {These solutions are the infinite-$Re$ analogues of the so-called ``unimodal'' stationary and time-periodic solutions of the Navier-Stokes equation at high $Re$ \citep{kim2010,gallet2013,kim2015,kim2017}.}
Unlike the vast majority of exact solutions {of 2D Euler} reported previously, many of the solutions reported here are dynamically relevant: they provide a reasonably accurate description of the spatial structure and dynamics of the large-scale component of high-Reynolds number turbulent flow. Hence they should be thought of as the proper exact coherent structures in the limit of infinite Reynolds number. The time-periodic solutions were found to be particularly important: large-scale turbulent flow was found to mirror this type of recurrent solutions of the Euler equation surprisingly frequently.

Another important result of this study is that recurrent solutions of Euler belong to families parameterized by an infinite number of continuous parameters. Some of these parameters correspond to translational and scaling symmetries of Euler, while the rest can be thought of as ``shape'' parameters describing the spatial profile and relative position of the vortices. Clearly, only a fraction of the members of those solution families are dynamically relevant. What distinguishes the region of the huge parameter space that is dynamically relevant from the rest is yet to be determined, although solvability conditions can provide useful insight.

Rather unexpectedly, recurrent solutions of Euler featuring a pair of counter-rotating vortices were found to be only weakly unstable. This is in contrast with the intuition generated by previous numerical studies of {weakly turbulent} flows which suggests that exact coherent structures become progressively more unstable as the Reynolds number increases. The likely reason this intuition breaks down is that the recurrent solutions of Euler investigated here are analogues of classical coherent structures describing large-scale flows. On the other hand, continuation of recurrent solutions describing weakly turbulent flows to higher Reynolds numbers typically yields coherent {\it substructures} describing small-scale flows. It is these substructures that become increasingly unstable. 

Another interesting fundamental question that remains unresolved is the relation between the spatial and temporal structure of the prevailing time-periodic exact coherent structures and the enstrophy flux towards small scales (direct cascade). The Kraichnan-Leith-Batchelor theory completely fails to describe the turbulent flow studied here, yet we still find a power-law scaling of the energy spectrum, albeit with a non-integer exponent. Subsequent work should determine why a power law emerges and what the exponent is.

\backsection[Acknowledgements]{This material is based upon work supported by the National Science Foundation under grant no. 2032657.}

\backsection[Declaration of interests]{The authors report no conflict of interest.}

\backsection[Supplementary data]{\label{SupMat}Movies corresponding to some figures can be found at \newline \url{https://cns.gatech.edu/roman/2DEuler/}.}

\backsection[Data availability statement]{Exact coherent structures and their key properties as well as matlab codes for solving the governing equations and Newton-GRMES solvers are available on GitHub at \url{https://github.com/cdggt/euler2D/tree/main}}

\bibliographystyle{jfm}
\bibliography{refs.bib}


{
\section*{Appendix A: Numerical methods}

\subsection*{Solving the Euler and Navier-Stokes equations}

Both the Euler equation and the Navier-Stokes equation can be written entirely in terms of the stream function $\psi=-\nabla^{-2}\omega$,
\begin{align}
    \partial_t\psi=g(\psi),
\end{align}
with an appropriately defined function $g(\psi)$. In the case of the Navier-Stokes equation,  
\eb g(\psi) = \nabla^{-2}\left(\pd{\psi}{x}{}\pd{}{y}{}\nabla^2\psi-\pd{\psi}{y}{}\pd{}{x}{}\nabla^2\psi-\varphi({\bf x},{\bf y})\right)+\frac{1}{Re}\nabla^2\psi, \en
where $\varphi$ is the forcing profile introduced in Equation \eqref{eq:nse}.

To handle a wide range of time scales present in turbulence, time-stepping was performed using variable-time-step Runge-Kutta-Fehlberg scheme
\eb 
\psi(t+\Delta t) &= \psi(t) + \Delta t\left(\frac{47}{450}k_1+\frac{12}{25}k_3+\frac{32}{225}k_4+\frac{1}{30}k_5+\frac{6}{25}k_6\right), 
\en 
where
\eb 
k_1 &= g\left(\psi(t)\right) , \\
k_2 &= g\left(\psi(t)+\frac{2\Delta t}{9}k_1\right) , \\
k_3 &= g\left(\psi(t)+\frac{1\Delta t}{12}k_1+\frac{1\Delta t}{4}k_2\right) ,\\
k_4 &= g\left(\psi(t)+\frac{69\Delta t}{128}k_1-\frac{243\Delta t}{128}k_2+\frac{135\Delta t}{64}k_3\right) , \\
k_5 &= g\left(\psi(t)-\frac{17\Delta t}{12}k_1+\frac{27\Delta t}{4}k_2-\frac{27\Delta t}{5}k_3+\frac{16\Delta t}{15}k_4\right) , \\
k_6 &= g\left(\psi(t)+\frac{65\Delta t}{432}k_1-\frac{5\Delta t}{16}k_2+\frac{13\Delta t}{16}k_3+\frac{4\Delta t}{27}k_4+\frac{5\Delta t}{144}k_4\right) , 
\en 
The step size $\Delta t$ was adjusted according to 
\eb \Delta t_{\rm new} =
\epsilon \mathcal E^{-1/5}\Delta t_{\rm old}
\en 
with fixed tolerance $\epsilon$ (typically $\epsilon=1.5$) and 
\eb \mathcal E = \Delta t \left\|\frac{1}{360}k_1-\frac{128}{4275}k_3-\frac{2197}{75240}k_4+\frac{1}{50}k_5+\frac{2}{55}k_6 \right\|. 
\en

In the case of the Euler equation, where 
\eb g(\psi) = \nabla^{-2}\left(\pd{\psi}{x}{}\pd{}{y}{}\nabla^2\psi-\pd{\psi}{y}{}\pd{}{x}{}\nabla^2\psi\right), \en 
time-dependent solutions were computed using a fixed-step-size fourth-order Runge-Kutta method:
\eb \psi(t+\Delta t) &= \psi(t)+\Delta t\left(\frac16k_1+\frac13k_2+\frac13k_3+\frac16k_4\right),
\en 
where
\eb
k_1 &= g\left(\psi(t)\right), \\
k_2 &= g\left(\psi(t)+\frac{\Delta t}{2}k_1\right), \\ 
k_3 &= g\left(\psi(t)+\frac{\Delta t}{2}k_2\right), \\
k_4 &= g\left(\psi(t)+\Delta t k_3\right). \\
\en 
To prevent contamination of smooth solutions by high-frequency noise associated with discretization and truncation errors, we added a term $\hat{D}\psi$ representing hyperviscosity to $g(\psi)$, where 
\eb\label{eq:hv} 
\hat D e^{i{\bf k}\cdot{\bf x}} = \begin{cases} 
\alpha(k_v-k)\exp\left(\frac{\beta}{k_v-k} \right) e^{i{\bf k}\cdot{\bf x}},\quad&\text{if } k>k_v,\\
0,&\text{otherwise}.
\end{cases}\en
We used the following choices of parameters: $k_v=40$, $\alpha=0.5$ and $\beta=1/13$. We also checked that converged solutions change negligibly if $k_v$ is increased to 64. The choice of the functional form of the hyperviscous term is arbitrary, so long as it has no effect on wavenumbers below the threshold $k_v$, while the high wavenumbers are strongly suppressed. For our choice of $k_v$, hyperviscosity only affects the high frequencies that are essentially absent in large-scale flows.

\begin{figure}
    \centering
    \subfloat[]{\includegraphics[width=0.45\textwidth]{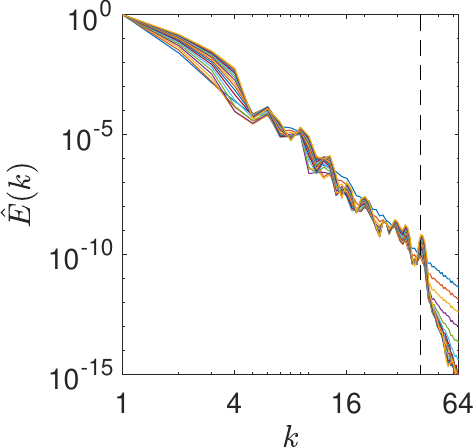}}\hspace{10mm}
    \subfloat[]{\includegraphics[width=0.45\textwidth]{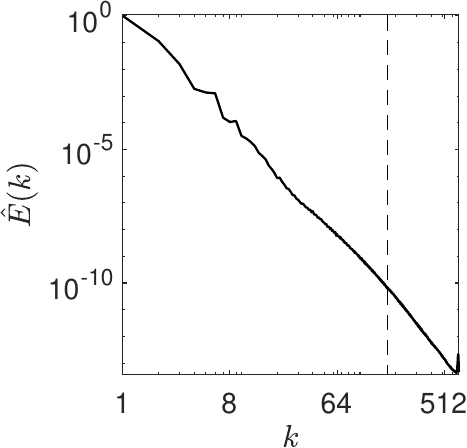}}
    \caption{Time-averaged energy spectra of (a) the periodic orbits in the family presented in Figure \ref{f:pofam1} computed on a $256\times 256$ grid and (b) turbulent trajectory computed on a $2048\times2048$ grid. The dashed vertical line in (a) represents the hyperviscosity threshold $k_v$, and in (b) it represents the two-thirds dealiasing cutoff corresponding to a $512\times512$ grid. The spectra at $k=1$ have been normalized to unity in both cases. }
    \label{f:spectra}
\end{figure}

Spatial derivatives were all computed spectrally for both the Euler and Navier-Stokes equation. 
Additionally, a $2/3$ dealiasing scheme was implemented for numerical stability.
Typically, we computed solutions to the Navier-Stokes equation on a $512\times512$ grid, although we also checked the key results for consistency using grids up $2048\times2048$. A time-averaged energy spectrum shown in Figure \ref{f:spectra} indicates that, for a $2048\times2048$ grid, the energy decreases by ten orders of magnitude (corresponding to an $O(10^{-5})$ relative error for the velocity field) between modes with $k=1$ and modes with $k=170$ (indicated by the dashed line in the Figure), the latter corresponding to the dealiasing cutoff on a $512\times512$ grid. This is consistent with the result shown in Figure \ref{f:spectrum}(a) and suggests that our simulations of turbulent flow are fairly well-resolved even on $512\times512$ grids. 

Solutions of the Euler equation were computed on a $256\times256$ grid, which provides sufficient resolution for the smooth solutions we are interested in. Indeed, as seen in Figure \ref{f:spectra}, for a family of time-periodic ECSs, we find the time-averaged energy to decrease by ten orders of magnitude between modes with $k=1$ and modes with $k=k_v=40$ (indicated by the dashed line in the Figure), suggesting that hyperviscosity has a negligible effect on the solution. Note that dealiasing cutoff corresponds to a far higher frequency $k=85$ in this case. 
\subsection*{Newton-GMRES solver}

ECSs of all types were computed using a Krylov subspace Newton method (see, e.g. \citep{Viswanath2007}), which uses GMRES to solve the equation 
\eb\label{eq:Newton}
J(\omega_j) \Delta\omega_j=F(\omega_j)
\en
for the vorticity field, where $F$ represents the relevant conditions for either equilibria or periodic orbits (specified below), and $J$ is the Jacobian of $F$. The initial condition $\omega_0$ is either taken to be a properly smoothed snapshot of the turbulent flow field or constructed using homotopy from previously converged solutions.
Vorticity is then updated iteratively as $\omega_{j+1}=\omega_j-\Delta\omega_j$ until the relative error $\zeta=\|F(\omega)\|/\|\omega\|$ becomes less than $5\times10^{-5}$.
Note that, in terms of the velocity field, the accuracy of our converged solutions corresponds to an $O(10^{-5})$ relative residual, which is the same as the relative error of our numerical simulations. Hence, converged solutions can be considered numerically exact, with their accuracy limited mainly by the spatial and temporal resolution of the simulations. 

Since ECSs of the Euler equation have marginal directions, $J$ becomes non-invertible near a solution of $F(\omega)=0$. Although, due to the presence of marginal directions, equation \eqref{eq:Newton} has infinitely many solutions in the full state space, its projection
\eb
\hat{P}_KJ(\omega_j) \Delta\omega_j=\hat{P}_KF(\omega_j)
\en
onto the Krylov subspace 
\eb  K = {\rm span}\{Jx, J^2x, J^3 x,\cdots\}. \en 
constructed using Arnoldi iteration (where vector $x$ does not lie in the kernel of $J$) has a well-defined unique solution so long as $\dim K\leq \dim \text{image}(J)$. In practice, the latter condition is always satisfied.

\subsection*{Equilibria}

For equilibria, we used the function
\eb F(\omega,{\bf v}) = \left(({\bf u-v})\cdot\nabla\omega,\,E - E_0,\, \phi_x ,\, \phi_y  \right), \en 
where the additional constraints fix the energy $E$ and the spatial position $(\phi_x,\phi_y)$ of the solution. The energy $E_0$ was typically set equal to that of the initial condition $\omega_0$.
The velocity $\bf v$ of the reference frame was set to be zero for absolute equilibria and treated as part of our solution space for relative equilibria. 

\subsection*{Periodic orbits}

For time-periodic solutions, we used the objective function 
\eb\label{eq:POobj} F(\omega) = \omega-\hat L\hat\Phi_T\omega, \en 
where $\hat\Phi_T$ denotes the time-$T$ map, i.e., $\hat{\Phi}\omega(0)=\omega(T)$, and $\hat L$ is a linear operator that includes translations (to allow relative solutions) and discrete rotations (to allow preperiodic solutions such as those presented in Figure \ref{f:eqfam1}). For time-periodic solutions, we fixed the period $T$ instead of the energy, since the former constraint was more straightforward to implement numerically.

The introduction of the hyper-viscous term \eqref{eq:hv} was crucial for converging time-periodic orbits, as it prevented Newton iterations from introducing high frequencies into the solution. As the solution become better converged, the threshold frequency $k_v$ can be increased to make sure the converged solution satisfies the Euler equation to a higher accuracy. 

\subsection*{Homotopy}

Solution families reported here have been constructed using homotopy. In its standard form, homotopy represents a linear interpolation 
\begin{align}
\label{eq:homotopy1}
    {\bf u}(\sigma)=\sigma {\bf u}(0)+(1-\sigma){\bf u}(1)
\end{align}
between a pair of states ${\bf u}(0)$ and ${\bf u}(1)$ characterized by a continuous parameter $\sigma\in[0,1]$. Note that the flow fields defined by equation \eqref{eq:homotopy1} generally do not satisfy the Euler equation even when both ${\bf u}(0)$ and ${\bf u}(1)$ do. Hence, one could use these flow fields as initial conditions to the Newton-GMRES solver. This naive approach works sufficiently well for solutions that are close to each other.

For solutions that are not particularly close, simple interpolation generates initial conditions that are not close to solutions either unless $\sigma$ is close to 0 or 1. In such cases, a sequential interpolation scheme can be used which employs a homotopy between previously converged solutions. Namely, for some discrete set $\{\sigma_j\}\in(0,1)$ and previously converged solutions ${\bf u}(\sigma_{j-1})$ and ${\bf u}(1)$ we define the initial condition for ${\bf u}(\sigma_{j-1})$ as 
\eb {\bf u}^*(\sigma_j) = \frac{(1-\sigma_j){\bf u}(\sigma_{j-1})+(\sigma_j-\sigma_{j-1}){\bf u}(1)}{1-\sigma_{j-1}}.
\en 
This method yields better (smoother) approximations to continuous families connecting distant states. Indeed, while the homotopy may yield a smooth family of initial conditions ${\bf u}^*(\sigma_j)$, converged states ${\bf u}(\sigma_j)$ may not form a smooth family since the difference ${\bf u}^*(\sigma_j)-{\bf u}(\sigma_j)$ is not guaranteed to be small unless the initial condition is close to a solution. Most of the families of solutions described in the paper were constructed via this method. 
}

\subsection*{Smoothing of initial conditions}

\begin{figure}
    \centering
        \subfloat[]{\includegraphics[height=0.33\textwidth]{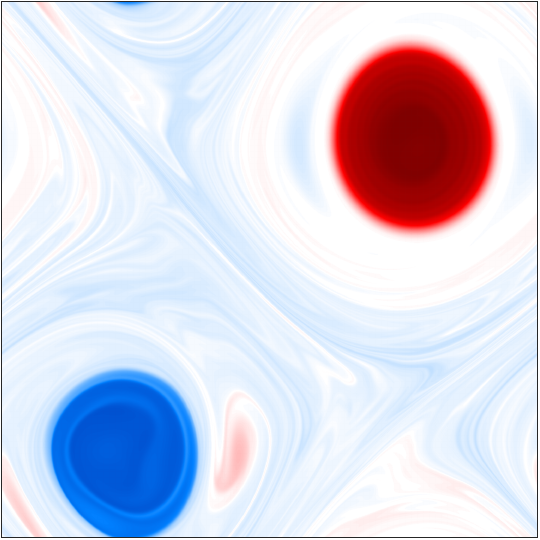}}
       \hspace{10mm}
       \subfloat[]{\includegraphics[height=0.33\textwidth]{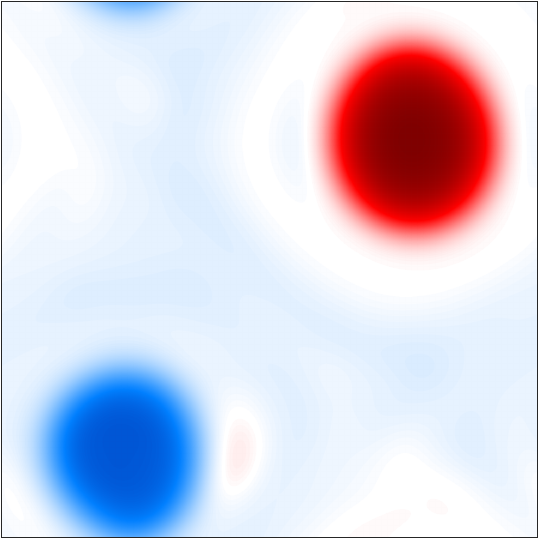}}\\
       \subfloat[]{\includegraphics[height=0.33\textwidth]{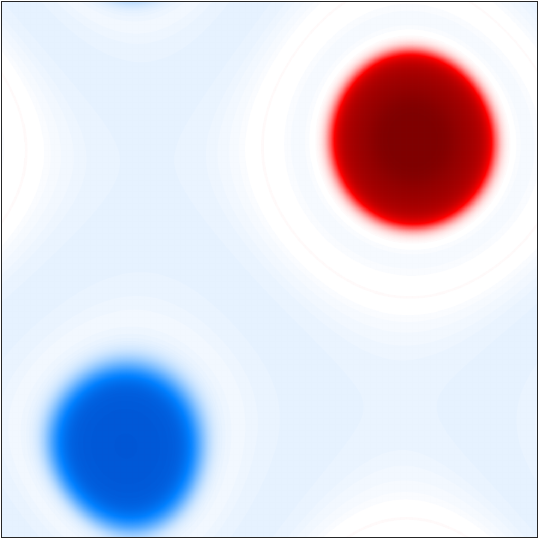}}
       \hspace{10mm}
       \subfloat[]{\includegraphics[height=0.33\textwidth]{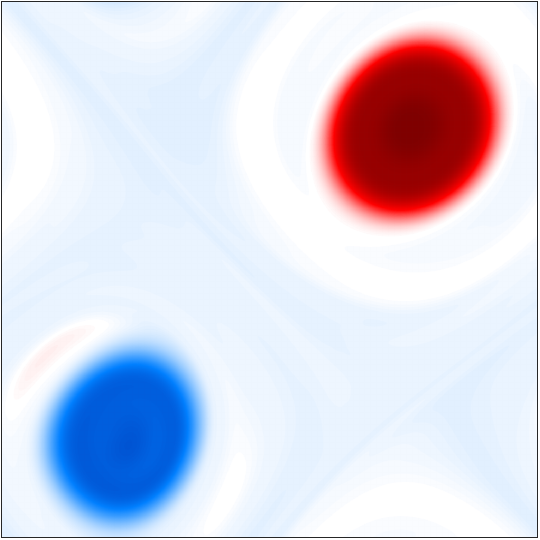}}
    \caption{A snapshot of turbulent flow field (a) and initial conditions for the Newton-GMRES solver obtained by applying spectral smoothing with $k_s=7$ (b), stream function smoothing with $\sigma=0.5$ (c), and hyperviscous smoothing with $k_v=7$, $\alpha=1$, and $\beta=1/13$. The relative error $\zeta$
    for the corresponding initial conditions is $0.226$ in (a), $0.0932$ in (b), $0.105$ in(c), and $0.163$ in (d).}
    \label{f:blurs}
\end{figure}

In order to speed up convergence of Newton-GMRES iterations and improve the success rate of the solver, initial conditions found from recurrence analysis of turbulent flow should be properly smoothed. The rate of success depends rather sensitively on the choice of both the smoothing method and its various parameters. While quantifying the rate of success and convergence speed as a function of all these choices is outside the scope of our study, we note that sufficiently aggressive smoothing yields an effectively 100\% success rate, as is the case for the solutions shown in Figure \ref{f:TrajectoryLSECS}. 
Below we describe three different smoothing methods that have been used to prepare initial conditions for the Newton-GMRES solver, with the results compared in Figure \ref{f:blurs} for an initial condition that corresponds to a nearly-time-periodic large-scale flow.

{\bf Spectral smoothing:} The flow field can be smoothed by applying {\ an axially symmetric} mask in Fourier space. The most common {\  and effective} choice was the Gaussian blur which corresponds to the mask
\eb {\  M({\bf k};k_s)} = \exp\left(-\frac{k^2}{k_s^2}\right) 
\en 
with a characteristic wavenumber $k_s$ controlling the degree of smoothing.
Another mask that was occasionally used is 
\eb  {\  M({\bf k};k_s)} = \min\left(1,\frac{k_s}{k} \right). 
\en 
{\  In a few instances we also used the Fourier filtering operator $\hat{L}_{k_s}$ used to split the flow into the large- and small-scale component, which corresponds to the mask
\eb
M({\bf k};k_s) = \frac12\left[1-\tanh\left(\frac{k-k_s}\ell\right)\right],
\en
where we set $\ell=5$.

Spectral smoothing is a very standard and simple approach which is effective at removing small-scale structure while preserving some large-scale structure of the flow. It can noticeably change the dynamics, however. }

{\bf Hyperviscous smoothing:} The fine structures in the flow field can also be effectively eliminated via the action of hyperviscosity over a suitable long period of time. For time-periodic states, we evolve the flow over one period using the time-integrator for the Euler equation with $k_v$ set to a relatively low value (compared with that used in the Newton-GMRES solver).  {\  This approach allows small-scale structure to be removed without significantly affecting the spatial or temporal structure of the large-scale flow.}

{\bf Stream function smoothing:} Solutions of the Euler equation can vary quickly transverse to, but not along, the streamlines of the flow. Hence it is helpful to average the vorticity field $\omega$ along the instantaneous streamlines. The smoothed vorticity field is then given by 
\eb 
\bar\omega(x,y) = \iint \exp\left(\frac{\left[\psi(x,y)-\psi(x',y')\right]^2}{2\sigma^2}\right)\omega(x',y')\,dx'dy'.
\en 
{\  Unlike the other two smoothing approaches, this method is better at preserving the edges of the vortex cores where vorticity changes rapidly, while damping out asymmetrical structures around the vortices. For instance, the flow shown in panels (a), (b), and (d) of Figure \ref{f:blurs}, features vortices with pronounced azimuthal structure; this structure has been largely removed by the stream function smoothing as shown in panel (c). Hence, this smoothing method is unsuitable for computing solutions featuring structured vortices, such as the tripolar vortex seen in Figure \ref{f:tripole1}.
}

In particular, in computing the solutions reported in Figure \ref{f:TrajectoryLSECS}, we created initial conditions by sequentially performing stream function smoothing followed by spectral smoothing and, ultimately, hyperviscous smoothing, with the combination of three methods reducing the relative error $\zeta$ from $O(0.1)$ to $O(10^{-3})$. A few tens of Newton iterations were then sufficient to achieve convergence with $\zeta<5\times 10^{-5}$ for all initial conditions.  {\ This aggressive smoothing is the reason why some properties of converged solutions differ substantially from those of the corresponding turbulent flow snapshots, especially at later times.}

\section*{Supplementary Material}

Movies of turbulent flow and various recurrent solutions of the Euler equation are available at the URL \url{https://cns.gatech.edu/roman/2DEuler/}

\medskip\noindent
Movie S1: Fully-resolved turbulent flow (bottom left), the corresponding large-scale flow (bottom right) and the recurrence diagram (top). The black line shows the current time instant.

\medskip\noindent
Movie S2: Unstable periodic orbit with temporal period $T=10.02f$ a snapshot of which is shown in Figure 2(d).

\medskip\noindent
Movie S3: Unstable periodic orbit with the temporal period $T=1.045$ a snapshot of which is shown in Figure 10(c).

\medskip\noindent
Movie S4: A family of solutions connecting unstable periodic orbits shown in Figures 11(a) and 11(c).

\medskip\noindent
Movie S5: A family of solutions connecting the equilibrium shown in Figure 12(a) and the periodic orbit shown in Figure 12(b).

\end{document}